\RequirePackage{fix-cm}
\documentclass[smallcondensed]{svjour3}     
\smartqed  
\usepackage{graphicx}
\usepackage{xcolor}
\usepackage{physics}
\usepackage{subcaption}
\usepackage{amssymb}
\usepackage{mathtools}
\usepackage[title]{appendix}
\usepackage{lineno}
 
\newcommand{\hr}{\hat{r}}
\newcommand{\hz}{\hat{z}}
\newcommand{\hphi}{\hat{\phi}}
\newcommand{\hh}{\hat{h}}
\newcommand{\hp}{\hat{p}}

\newcommand{\tilr}{\tilde{r}}

\newcommand{\tz}{\tilde{z}}
\newcommand{\tilh}{\tilde{h}}
\newcommand{\tphi}{\tilde{\phi}}
\newcommand{\tp}{\tilde{p}}

\newcommand{\br}{\bar{r}}
\newcommand{\bz}{\bar{z}}
\newcommand{\bphi}{\bar{\phi}}
\newcommand{\bu}{\bar{u}} 

\newcommand{\bh}{\bar{h}}
\newcommand{\bp}{\bar{p}}
\journalname{Journal of Engineering Mathematics}

\begin{document}

\title{Droplet impact onto a spring-supported plate: analysis and simulations
}


\author{Michael J. Negus         \and
        Matthew R. Moore      \and
        James M. Oliver \and
        Radu Cimpeanu
}


\institute{Michael J. Negus \and Matthew R. Moore \and James M. Oliver \and Radu Cimpeanu \at
              Mathematical Institute, University of Oxford, Radcliffe Observatory Quarter, Oxford OX2 6GG, UK \\
              \email{michael.negus@maths.ox.ac.uk}           
           \and
           Radu Cimpeanu \at
            Mathematics Institute, Zeeman Building, University of Warwick, Coventry CV4 7AL, UK
            \email{Radu.Cimpeanu@warwick.ac.uk} 
}

\date{Received: date / Accepted: date}

\maketitle

\begin{abstract}
The high-speed impact of a droplet onto a flexible substrate is a highly nonlinear process of practical importance which poses formidable modelling challenges in the context of fluid-structure interaction. We present two approaches aimed at investigating the canonical system of a droplet impacting onto a rigid plate supported by a spring and a dashpot: matched asymptotic expansions and direct numerical simulation (DNS). In the former, we derive a generalisation of inviscid Wagner theory to approximate the flow behaviour during the early stages of the impact. In the latter, we perform detailed DNS designed to validate the analytical framework, as well as provide insight into later times beyond the reach of the proposed mathematical model. Drawing from both methods, we observe the strong influence that the mass of the plate, resistance of the dashpot and stiffness of the spring have on the motion of the solid, which undergoes forced damped oscillations. Furthermore, we examine how the plate motion affects the dynamics of the droplet, predominantly through altering its internal hydrodynamic pressure distribution. We build on the interplay between these techniques, demonstrating that a hybrid approach leads to improved model and computational development, as well as result interpretation, across multiple length- and time-scales.

\keywords{Impact \and Droplets \and Interfacial flows \and Asymptotic analysis \and Direct numerical simulation \and Fluid-structure interaction}
\end{abstract}

\section{Introduction}
\label{intro}

Droplet impacts are a rich and ubiquitous phenomenon in both nature and industry, from inkjet printing \cite{Hoath2016} to pesticide spray deposition \cite{Smith2000} and estimating the early stages of oily aerosol dispersal in the atmosphere after large-scale spills \cite{Murphy2015}. The dynamics of these processes are governed by the complex flow physics of the droplet and the surrounding gas, as well as the properties of the substrate, such as wettability \cite{Antonini2012} and roughness \cite{Ellis2011}. An additional layer of complexity is added to the system if the substrate is deformable, meaning that the force of the impactor causes the substrate to move or change shape. A common example of droplet impact onto deformable substrates is that of rainfall onto leaves \cite{Gilet2014}. Previous theoretical and experimental studies of this class of systems include droplet impact onto cantilever beams \cite{Gart2015}, silicone gels \cite{Howland2016a} and elastic membranes \cite{Pepper2008}.

Recent developments in experimental imaging techniques, improving in both frame rate and spatial resolution, have reinvigorated investigative efforts in high-speed impact \cite{Josserand2016}, revealing previously inaccessible features due to the small, rapidly developing regions upon impact, such as the impingement of micro-drops \cite{Visser2015} and early azimuthal instabilities of the ejecta \cite{Li2018}. Similarly, the increase in high performance computing resources have allowed increasingly efficient direct numerical simulations (DNS) of these systems to be performed \cite{Cimpeanu2018,Lopez-Herrera2019,Thoraval2012}, which are intensive due to the rapidly evolving interfaces, multi-scale flow features and high density and viscosity ratios present. These experimental and computational difficulties mean that rigorous mathematical modelling of such flows is of key importance to any comprehensive investigation, as analytical approximations to the flow can provide insight into the underlying physical processes, greatly enhance predictive capabilities in regimes otherwise difficult to examine, and save computational resources required for numerical simulations. In all cases, the deformability of the substrate makes studying these systems even more complex, from the difficulty in observing small deformations experimentally to the additional degrees of freedom required for numerical study.

Studies of droplet impact usually focus on a specific timescale, as examining the rapidly evolving interfaces makes universal investigations challenging. Shortly before impact, the cushioning effect of the gas layer between the droplet and the substrate leads to high pressures which cause the bottom of the droplet to dimple. This interfacial deformation results in a gas bubble being entrapped inside the droplet upon impact, which has been widely observed experimentally \cite{Thoroddsen2005}, as well as reproduced in numerical studies \cite{Thoraval2012}, and modelled analytically \cite{Hicks2010a}. At early stages of the impact, close to the points of contact, the free surface rapidly turns over and begins to spread across the substrate. During this timescale, instabilities in the free surface can cause micro-droplets to be rapidly expelled, which has been observed experimentally using high-speed photography \cite{Thoroddsen2012}. Numerical schemes with adaptive mesh refinement \cite{Popinet2003,Popinet2015} allow for the computational study of this early timescale by concentrating resources in the small region close to the substrate \cite{Cimpeanu2018,Philippi2016,Thoraval2012}. Analytical approaches adapt Wagner theory \cite{Howison2005,Howison1991,Scoland2001,Wagner1932}, an inviscid fluid model for solid-liquid impact inspired by the study of aircraft landing on water and ship slamming. Later in the impact process, once the droplet begins to fully spread across the substrate, viscosity and surface tension, as well as the chemistry of the substrate, tend to play a more significant role. Various physical parameters determine whether the droplet retracts, rebounds or splashes; see \cite{Josserand2016} for an extensive review on the experimental studies on late impact. Quantities of interest at this timescale include the minimum thickness and the maximum diameter of the droplet as it spreads, and in particular, how these depend on the physical properties of the liquid and the substrate have been the focus of previous theoretical studies \cite{Eggers2010,Wildeman2016}.

Experimental work focused on droplet impact onto elastic substrates is much less common. Recent investigations have examined the impact of droplets onto the end of cantilever beams \cite{Gart2015,Weisensee2016}, drawing close parallels to the impact onto leaves. In both cases, the impact of the droplet excited oscillations at the end of the beam, with characteristic time periods of the order of the late impact timescale. Different length beams were considered in order to show how stiffer beams resulted in oscillations with higher frequencies. Other studies concerning droplet impact onto elastic membranes \cite{Pepper2008} and silicone gels \cite{Howland2016a} show that the compliance of the substrate strongly affects the splashing threshold (regarded here as the minimum velocity necessary to observe splashing), which is significantly increased when the stiffness of the substrate is reduced. 

Fluid-structure interaction problems are notoriously difficult to study numerically  and allowing the substrate to deform only exacerbates this. If the substrate only exhibits translational motion (without accounting for bending), then the problem can be simplified by considering a moving frame of reference, centred on the substrate \cite{Li2002}, whereas substrates which exhibit bending need to be considered using more complex techniques such as the immersed boundary method \cite{Peskin2002}. Two-phase flow systems with immersed boundaries have not been studied extensively, with only a few noticeable exceptions \cite{Patel2017,Sun2016}. Despite considering complex moving boundaries (such as a twin screw kneader), the motion of the boundaries were prescribed rather than resulting from fluid-structure interaction, although the proposed methods could in principle be extended to consider this. More recently, the impact of droplets in capillary-dominated regimes onto a flexible substrate has been modelled using the Lattice-Boltzmann method \cite{Xiong2020}, focusing on the spreading and rebound of the droplets and how the contact time is affected by the bending stiffness. The late-time spreading and rebound dynamics of an undamped plate-spring system have recently been studied numerically \cite{Zhang2020}, inspired by the feathers of kingfishers. It was found that springs with certain stiffness values can shorten the length of time the droplet is in contact with the substrate, as well as increase the speed the droplet rebounds after impact.

Analytical models for a liquid impacting a deformable substrate (or vice-versa), on the other hand, have been proposed for over half a century. One of the earliest models investigated a droplet impacting onto an elastic half-space by imposing a constant uniform pressure over a circle whose radius increased in proportion to the square root of time \cite{Blowers1969}. The full hydroelastic problem of the impact of a two-dimensional wave onto an Euler-Bernoulli beam has previously been studied using Wagner theory \cite{Korobkin1998}, and more recently this analytical model has been extended to study the axisymmetric impact of a droplet onto an elastic plate, where the elastic plate has a radius much smaller than the droplet and its deflection governed by thin-plate theory \cite{Pegg2018}. A thorough parameter study was conducted for different types of plate, and regimes where the elasticity of the plate could cause splashing of the droplet at early times, defined as the detachment of the splash sheet from the surface of the plate, have been identified.

To the best of our knowledge, a comprehensive study considering the fluid-structure interaction between an impacting droplet and a compliant substrate that systematically compares accurate numerical results to analytical or experimental counterparts has yet to materialise. One of the simplest types of deformable substrate systems which exhibits both elastic and damping effects is a rigid plate suspended by a Hookean spring and a linear dashpot, where the force of the impacting droplet causes the spring to compress, with the dashpot damping the motion. Here we present both a new analytical model extended from Wagner theory, as well as a direct numerical simulation platform for a droplet impacting onto a rigid plate supported by a spring and a dashpot. The system is used as a validation testbed for the two approaches, as well as a framework for an extensive parametric study focused on the effect of surface compliance on the ensuing drop dynamics in this challenging regime. We observe systematically how the influence of the substrate properties (mass, spring stiffness and damping factor) affects the fluid-structure interaction, emphasising how the resulting motion of the plate substantially alters the pressure field of the droplet, and, in turn, the hydrodynamic force exerted onto the plate, revealing a rich coupling of the different forces at play in the system. 

The rest of this paper is structured as follows. We outline the system geometry and general mathematical framework, as well as discuss assumptions at the level of the analytical and numerical models in \textsection\ref{sec:formulation}. In \textsection \ref{sec:analytical} we present the analytical model, deriving the leading-order composite solution for the pressure along the plate and the resulting hydrodynamic force. We describe the computational setup required for the direct numerical simulations and the specific algorithm we use to model the fluid-structure interaction in \textsection \ref{sec:dns}. The solutions from the analytical and numerical models are presented in \textsection \ref{sec:results}, where they are compared for a range of parameters of the plate-spring-dashpot system. We conclude by summarising our results and discussing the implications of this study and possible future extensions in \textsection \ref{sec:conclusion}. 

\section{Problem formulation}
\label{sec:formulation}

\begin{figure}[ht]
    \centering
    \includegraphics[width=0.75\textwidth]{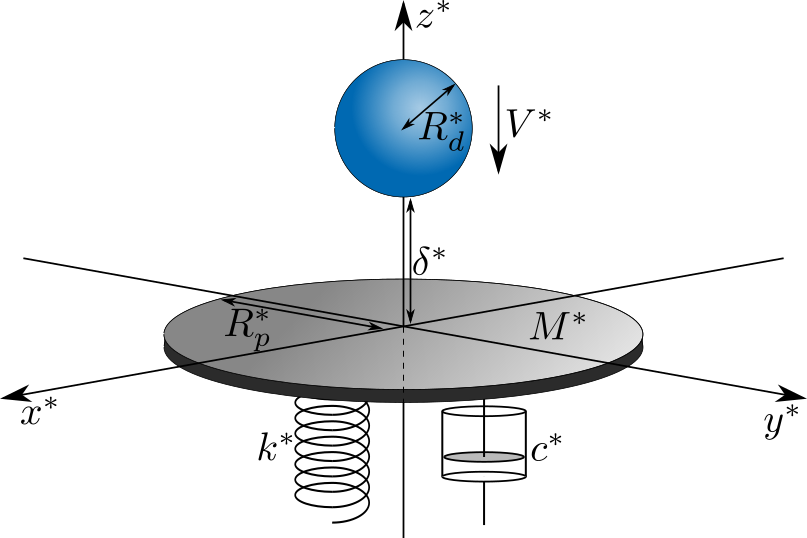}
    \caption{Schematic diagram of a spherical droplet of radius $R_d^*$ impacting with initial velocity $V^*$ onto a circular plate of radius $R^*_p$ and mass $M^*$. The plate is supported by a spring with spring constant $k^*$ and a dashpot with damping factor $c^*$.}
    \label{fig:dimensional_schematic}
\end{figure}

We consider the vertical impact of a droplet of incompressible, Newtonian liquid onto a rigid, planar, circular plate supported by a Hookean spring and linear dashpot. The droplet is initially spherical with radius $R_d^*$ and travelling uniformly downwards with speed $V^*$, surrounded by an incompressible gas. Throughout this study, dimensional quantities are denoted with a superscript $^*$.
The plate has radius $R^*_p$ and the plate-spring-dashpot system is initially in equilibrium. The bottom of the droplet is introduced at a height $\delta^* > 0$ above the plate. A Cartesian coordinate system $(x^*, y^*, z^*)$ is defined such that the surface of the plate lies in the $z^* = 0$ plane, the droplet falls along the $z^* > 0$ axis, and the bottom of the droplet is  given by $(x^*, y^*, z^*) = (0, 0, \delta^*)$ at the onset of the dynamics, as illustrated in Fig. \ref{fig:dimensional_schematic}.

The system is initialised at a time $t^* = t^*_0 = -\delta^* / V^*$. If the gas were absent, the plate would experience a zero net force until the droplet makes contact, and would therefore remain in equilibrium. In this case, the droplet would make contact with the plate at $t^* = 0$ with the plate stationary for $t^* < 0$. However the presence of the gas means there will be a pressure build-up prior to the impact \cite{Wilson1991,Hicks2010a}, resulting in a net force that causes the plate to accelerate downwards so the droplet will make contact at a time $t^*_c > 0$. 

The liquid comprising the droplet and the surrounding gas have  densities $\rho^*_l$, $\rho^*_g$ and viscosities $\mu^*_l$, $\mu^*_g$ respectively. The surface tension coefficient between the liquid and the gas is denoted by $\sigma^*$ (taken to be constant) and the acceleration due to gravity is $\vb{g}^* = g^* \vu{n}_z$, where $\vu{n}_z$ is the unit vector in the $z^*$ direction. The vertical position of the plate at time $t^*$ is $z^* = -s^*(t^*)$, where $s^*(t^*)$ is referred to as the plate displacement. 

Denoting the variables in the liquid and gas with a subscript $l$ and $g$ respectively, the Navier-Stokes equations are assumed to hold in each fluid, 
\begin{gather}
    \rho^*_i \left(\pdv{\vb{u}^*_i}{t^*} + (\vb{u}^*_i \vdot \grad) \vb{u}^*_i\right) = - \grad p^*_i + \mu^*_i \grad^2\vb{u}^*_i - \rho^*_i \vb{g}^*, \label{eq:dimen_momentum_eq}\\
     \div{\vb{u}^*_i} = 0, \label{eq:dimen_continuity_eq}
\end{gather}
where $i = l, g$, $\vb{u}^*_i$ is the velocity vector and $p^*_i$ represents the pressure in each fluid. 

The impermeability condition on the plate states that the fluid velocity must match the velocity of the plate along its surface,
\begin{equation}
\label{eq:dimen_impermeability}
    \vb{u}^*_i \vdot \vu{n}_z = - \dot{s}^*(t^*) \text{ for } z^* = - s^*(t^*), \; {x^*}^2 + {y^*}^2 < {R_p^*}^2,
\end{equation}
where the overdot denotes differentiation with respect to time. 

The kinematic condition at the multivalued free surface $z^* = h^*(x^*, y^*, t^*)$ is
\begin{equation}
\label{eq:dimen_free_surface_kinematic}
    v^*_n = \vb{u}^*_l \vdot \vu{n} \text{ on } z^* = h^*(x^*, y^*, t^*),
\end{equation}
where $\vu{n}$ is the unit outward-pointing normal vector to the free surface and $v^*_n$ is the normal speed of the free surface. Continuity of velocity and normal stress across the free surface are given by
\begin{equation}
\label{eq:dimen_free_surface_continuity}
    \vb{u}^*_g = \vb{u}^*_l, \quad \vu{n} \vdot [\mathcal{T}^*_g - \mathcal{T}^*_l] = - \sigma^* \kappa^* \vu{n} \text{ on } z^* = h^*(x^*, y^*, t^*),
\end{equation}
where $\mathcal{T}^*_i$ is the Cauchy stress tensor in each fluid and $\kappa^* = - \div{\vu{n}}$ is the curvature of the free surface. 

Initially, at $t^* = t^*_0$, the liquid has a uniform downwards velocity $V^*$,
\begin{equation}
    \vb{u}^*_l \equiv - V^* \vu{n}_z, 
\end{equation}
while the centre of the droplet is initially at $z^* = \delta^* + R_d^*$, meaning the free surface $h^*(x^*, y^*, t^*_0)$ satisfies
\begin{equation}
    {x^*}^2 + {y^*}^2 + (h^*(x^*, y^*, t^*_0) - \delta^* - R_d^*)^2 = {R_d^*}^2.
\end{equation}
The pressure $p^*_i$ is initially hydrostatic in each fluid, equal to $p^*_i = p^*_\text{atm} - \rho^*_i g^* z^*$, where $p^*_\text{atm}$ denotes atmospheric pressure. The gas far from the impact remains undisturbed for all times, so that
\begin{equation}
\label{eq:gas_far_field}
    \vb{u}^*_g \sim \vb{0}, \; p^*_g \sim p^*_\text{atm} - \rho^*_g g^* z^* \text{ as } {x^*}^2 + {y^*}^2 + {z^*}^2 \to \infty.
\end{equation}

The plate, which has total mass $M^*$, is supported by a Hookean spring with spring constant $k^*$, and a dashpot with damping factor $c^*$. At $t^* = t^*_0$, the plate is in equilibrium. Hence by Newton's third law, the force due to the compression of the spring balances the weight of the plate. Denoting the net hydrodynamic force applied to the plate in the downwards direction $-\vu{n}_z$ as $F^*(t^*)$, the displacement of the plate from this equilibrium is governed by
\begin{equation}
\label{eq:cantilever_ODE}
    M^* \ddot{s}^*(t^*) = F^*(t^*) - c^* \dot{s}^*(t^*) - k^* s^*(t^*).
\end{equation}

The net hydrodynamic force $F^*(t^*)$ is equal to the sum of the contributions from the hydrodynamic pressure and viscous stress above and below the plate. Assuming that the gas below the plate is at a constant pressure $p^*_\text{atm}$ and exerts a negligible amount of force due to viscous stress, then
\begin{equation}
\label{eq:dimen_hydrodynamic_force}
    F^*(t^*) = \iint\limits_{\substack{\sqrt{{x^*}^2 + {y^*}^2} \leq R^*_p \\z^* = 0^+}}(p^* - p^*_\text{atm}) - 2 \mu^*  \pdv{u_z^*}{z^*} \dd{x^*} \dd{y^*},
\end{equation}
where $p^* = p^*_l$, $\mu^* = \mu^*_l$, $u^*_z = u^*_{l, z}$ where the plate is wetted and $p^* = p^*_g$, $\mu^* = \mu^*_g$, $u^*_z = u^*_{g, z}$ where the plate is unwetted. 

\subsection{Non-dimensionalisation}
\label{subsec:non_dimensionalisation}
We take the initial droplet radius, $R_d^*$, and speed, $V^*$, as the characteristic length and velocity scales respectively. Then, choosing the advective and inertial time and pressure scales, we non-dimensionalise by setting
\begin{equation}
\begin{gathered}
	t^* = \frac{R_d^*}{V^*} t,\quad (x^*,y^*,z^*,h^*,s^*,R_p^*) = R_d^*(x,y,z,h,s,R_p), \\
	 \vb{u}^* = V^* \vb{u}, \quad p^* = p^*_\text{atm} + \rho^*_l {V^*}^2 p, \quad F^*(t^*)  = \rho^*_l {V^*}^2 {R_d^*}^2 F(t), 
\end{gathered}
\end{equation}
where $R_p$ is referred to as the dimensionless plate radius.

Under these scalings, the plate dispacement equation \eqref{eq:cantilever_ODE} becomes
\begin{equation}
\label{eq:dimensionless_plate_ode}
    \alpha \ddot{s}(t) + \beta \dot{s}(t) + \gamma s(t) = F(t),
\end{equation}
where
\begin{equation}
\label{eq:alpha_beta_gamma_def}
    \alpha = \frac{M^*}{\rho^*_l {R^*_d}^3}, \quad \beta = \frac{c^*}{\rho^*_l V^* {R_d^*}^2}, \quad \gamma = \frac{k^*}{\rho^*_l {V^*}^2 R_d^*}.
\end{equation}
The ratio between the mass of the plate and the mass of the droplet is equal to $3 \alpha / 4\pi$, hence $\alpha$ is referred to as the mass ratio. The damping factor, $\beta$, measures the strength of the resistance to motion due to the damping from the dashpot, and the stiffness factor, $\gamma$, measures the strength of the restoring force due to elastic compression of the spring. 

The relevant dimensionless parameters describing the flow dynamics are the Reynolds, Weber and Froude numbers, defined respectively by
\begin{equation}
	\text{Re} = \frac{\rho^*_l R_d^* V^*}{\mu^*_l}, \quad \text{We} = \frac{\rho^*_l R_d^* {V^*}^2}{\sigma^*}, \quad \text{Fr}^2 = \frac{{V^*}^2}{g^* R_d^*}.
\end{equation}
Finally, the ratios between the densities and the viscosities of the gas and the liquid are given by
\begin{equation}
	\rho_R = \frac{\rho^*_g}{\rho^*_l}, \quad \mu_R = \frac{\mu^*_g}{\mu^*_l}.
\end{equation}

\subsection{Modelling assumptions}
\label{subsec:assumptions}
In both \textsection \ref{sec:analytical} and \textsection \ref{sec:dns}, scenarios where the inertial effects of the impact are more significant than the effects of viscosity, surface tension and gravity are considered. Hence we assume throughout that the values of Re, We and Fr$^2$ are large. As an illustrative example, consider the impact of a droplet of water with radius $R_d^* = 1$ mm and velocity $V^* = 5$ m/s, surrounded by air under atmospheric conditions. This gives
\begin{equation}
\label{eq:dimensionless_numbers}
	\text{Re} \approx 4990, \quad \text{We} \approx 342, \quad \text{Fr}^2 \approx 2550,
\end{equation}
and we make the assumption that they remain large throughout the early stages of impact. Furthermore, the density and viscosity ratios for the air-water scenario are
\begin{equation}
\label{eq:ratios}
	\rho_R \approx 1.20 \times 10^{-3}, \quad \mu_R \approx 1.83 \times 10^{-2},
\end{equation}
which provides support for neglecting the effects of the gas phase in the analytical model (but not in the DNS). 

As the system is initially radially symmetric about the $z-$axis, we assume this symmetry remains in place throughout the impact and consider an axisymmetric coordinate system $(r, z)$, where $r^2 = x^2 + y^2$. This assumption restricts the applicability of the model away from systems that involve fully three-dimensional effects, such as prompt splashing \cite{Thoroddsen2012} and azimuthal instabilities of the ejecta sheet \cite{Li2018}. 

\section{Analytical model}
\label{sec:analytical}
The analytical model focuses on the dynamics shortly after the time of impact. The model follows the structure of previous works on axisymmetric Wagner theory, and the reader is directed to \cite{Howison2005,Howison1991,Moore2014,Oliver2002,Philippi2016} for more details on the general methodology for impact involving stationary substrates. In this context the droplet impacts the plate at $t = 0$, with the displacement and velocity of the plate equal to zero at $t = 0$. As the gas phase is ignored, all expressed quantities are in the liquid, and the subscript $l$ is dropped for brevity.

\subsection{Governing equations}
\label{subsec:asymptotic_governing}
Under these assumptions, the flow is irrotational for $t < 0$ and hence by Kelvin's circulation theorem will remain irrotational for $t > 0$. Therefore a velocity potential $\phi$ can be introduced, such that $\vb{u} = \grad{\phi}$. The dimensional continuity equation \eqref{eq:dimen_continuity_eq} transforms to Laplace's equation for $\phi$, 
\begin{equation}
\label{eq:laplace}
    \laplacian{\phi} = \pdv[2]{\phi}{r} + \frac{1}{r} \pdv{\phi}{r} + \pdv[2]{\phi}{z} = 0,
\end{equation}
and the dimensional momentum equation \eqref{eq:dimen_momentum_eq} results in the unsteady Bernoulli equation for $\phi$ and $p$,
\begin{equation}
\label{eq:bernoulli}
    p + \pdv{\phi}{t} + \frac{1}{2}|\grad{\phi}|^2 = C(t),
\end{equation}
for some $C(t)$. The absence of viscosity and surface tension means the continuity of normal stress boundary condition \eqref{eq:dimen_free_surface_continuity} reduces to specifying that $p = 0$ at the free surface. Finally, by neglecting the gas phase and liquid viscosity, the net hydrodynamic force \eqref{eq:dimen_hydrodynamic_force} is just equal to the integral of the pressure $p$ across the wetted part of the plate. Hence the governing equations are a set of non-linear equations for the velocity potential $\phi$, pressure $p$, free surface location $h$ and plate displacement $s$.

\subsection{Asymptotic structure}
Following the structure of previous analytical models for droplet impact \cite{Howison2005}, we identify that for $t \ll 1$, the radial extent of the penetration region (where the droplet would be below the $r$ axis were the plate not present) is $O(\sqrt{t})$. Given this, we introduce an arbitrarily small parameter $0 < \epsilon \ll 1$ and rescale time
\begin{equation}
    t = \epsilon^2 \hat{t},
\end{equation}
where $\hat{t} = O(1)$ as $\epsilon \to 0$. With the plate present, the free surface is violently displaced and the liquid is ejected along the plate in a splash sheet. The curve at which the free surface is vertical is called the \emph{turnover curve}, and for small times we assume its radial extent is close to that of the penetration region, meaning that
\begin{equation}
    r = \epsilon \hat{d}(\hat{t}), \quad \hat{d}(0) = 0,
\end{equation}
where the turnover curve $\hat{d}(\hat{t}) = O(1)$ as $\epsilon \to 0$. The bottom of the penetration region is at $z = - \epsilon^2 \hat{t}$, so if we assume that the plate acts to decelerate the vertical motion of the droplet at early times, then the plate position $z = - s(t) > - \epsilon^2 \hat{t}$, which motivates the rescaling 
\begin{equation}
    s(t) = \epsilon^2 \hat{s}(\hat{t}),
\end{equation}
where $\hat{s}(\hat{t}) = O(1)$ as $\epsilon \to 0$. For brevity, the hat notation is dropped for the rest of \textsection \ref{sec:analytical}. 

\begin{figure}
    \centering
    \includegraphics[width=0.8\textwidth]{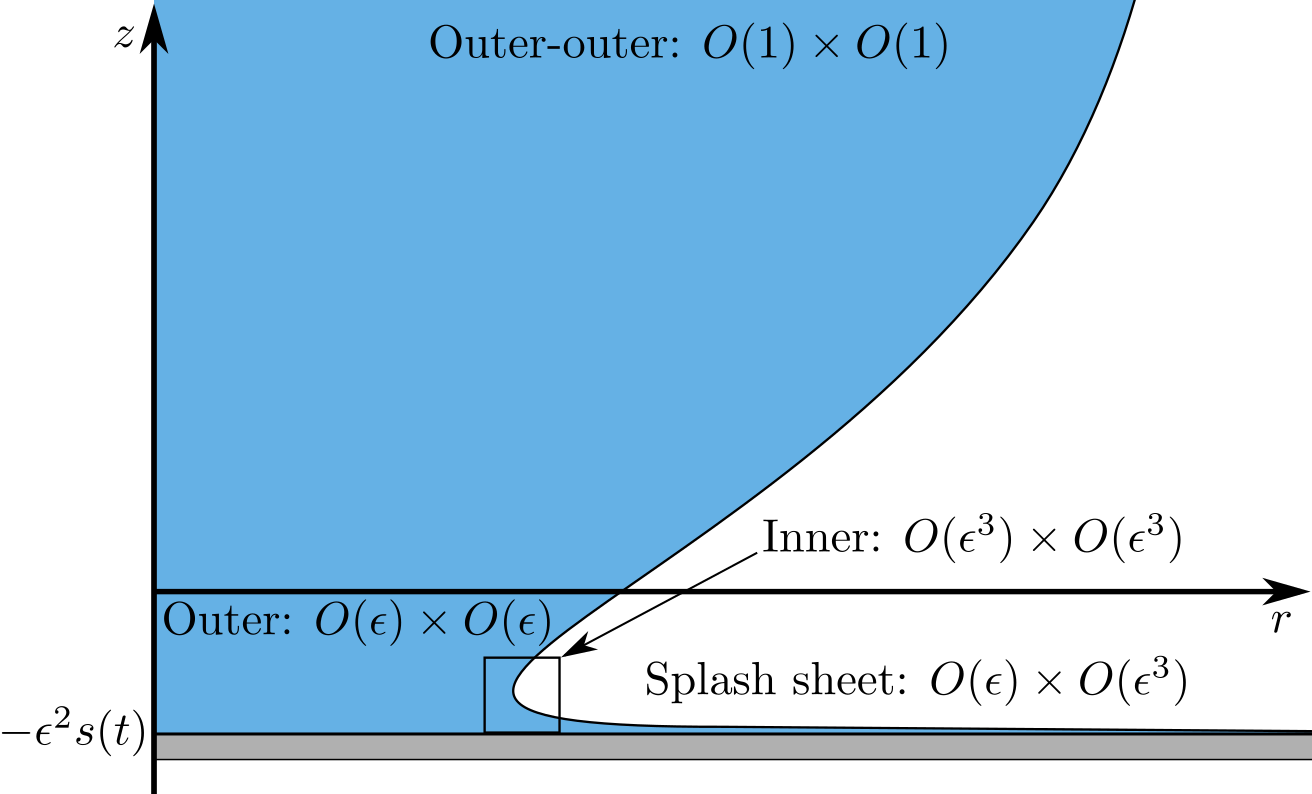}
    \caption{Schematic of the asymptotic structure of the system. The displacement of the plate $-\epsilon^2 s(t)$ has been exaggerated for visibility.}
    \label{fig:asymptotic_schematic}
\end{figure}

As $\epsilon \to 0$, the problem breaks down into four distinct spatial regions, as depicted in Fig. \ref{fig:asymptotic_schematic}. In the outer-outer region, for which $(r, z) = O(1) \times O(1)$, the bulk of the droplet is unaffected by the plate to leading-order, and is hence spherical and moving downwards with unit speed. This means that $C(t) = 1/2$ to leading order in the Bernoulli equation (\ref{eq:bernoulli}), and we shall not need to consider its higher-order corrections in the present analysis. The $O(\epsilon) \times O(\epsilon)$ region close to the centre of the plate is referred to as the outer region, and here the splash sheet is neglected and the boundary conditions are linearised onto the undisturbed plate location. This solution breaks down close to the turnover curve where the velocity and pressure become singular, which is corrected by introducing an inner region of size $O(\epsilon^3) \times O(\epsilon^3)$ in which the free surface turns over, ejecting fluid into the splash sheet. The thin, fast-moving splash sheet region is of size $O(\epsilon) \times O(\epsilon^3)$, emanating across the surface of the plate from the inner region. In this analysis, we assume the splash jet does not detach from the plate.

In the present analysis we shall consider the outer, inner and splash sheet regions in detail, however we forgo an analysis for the outer-outer region as it does not contribute to the leading-order hydrodynamic force on the plate. 

\subsection{Outer region}
\label{subsec:outer_region}
Guided by a well-known scaling argument \cite{Howison2005}, in the outer region we set
\begin{equation*}
    r = \epsilon \hr, \quad z = \epsilon \hz, \quad \phi = \epsilon \hphi, \quad h = \epsilon^2 \hh, \quad p = \frac{1}{\epsilon} \hp,
\end{equation*}
and expand $(\hphi, \, \hh, \, \hp, \, d, \, s) = (\hphi_0, \, \hh_0, \, \hp_0, \, d_0, \, s_0) + o(1)$ as $\epsilon \to 0$. The resulting governing equations in the outer region are
\begin{linenomath}
\begin{align}
    \pdv[2]{\hphi_0}{\hr} + \frac{1}{\hr} \pdv{\hphi_0}{\hr} + \pdv[2]{\hphi_0}{\hz} &= 0 &\text{ for } \hz > 0,& \label{eq:outer_laplace}\\
    \pdv{\hphi_0}{\hz} &= - \dot{s}_0(t) &\text{ on } \hz = 0,& \; \hr < d_0(t)  \label{eq:outer_plate_kinematic}\\
    \pdv{\hphi_0}{\hz} &= \pdv{\hh_0}{t}, &\text{ on } \hz = 0,& \; \hr > d_0(t), \label{eq:outer_free_surface_kinematic}\\
    \hphi_0 &= 0 &\text{ on } \hz = 0, & \; \hr > d_0(t), \label{eq:outer_free_surface_dynamic}
\end{align}
\end{linenomath}
where \eqref{eq:outer_laplace} is Laplace's equation for $\hphi_0$, \eqref{eq:outer_plate_kinematic} is the kinematic boundary condition on the plate, \eqref{eq:outer_free_surface_kinematic} is the kinematic boundary condition at the free surface and \eqref{eq:outer_free_surface_dynamic} is the dynamic boundary condition at the free surface, where \eqref{eq:outer_free_surface_dynamic} is found by integrating the leading-order Bernoulli equation \eqref{eq:bernoulli} with respect to time and applying the initial condition for $\hphi_0$. 

The far-field conditions as we tend towards the outer-outer region state that to leading-order, the flow is travelling downwards with speed 1 and the free surface is parabolic in $\hr$, such that
\begin{equation}
\label{eq:outer_far_field}
     \hphi_0(\hr, \hz, t) \sim - \hz, \text{ as } \sqrt{\hr^2 + \hz^2} \to \infty \; \text{ and } \hh_0(\hr, t) \sim \frac{1}{2} \hr^2 - t \text{ as } \hr \to \infty,
\end{equation}
with subsequent initial conditions
\begin{equation}
\label{eq:outer_initial_condition}
    \hh_0(\hr, 0) = \frac{1}{2} \hr^2, \quad s_0(0) = \dot{s}_0(0) = 0, \quad d_0(0) = 0.
\end{equation}

In addition, the so-called Wagner condition is needed in order to match to the inner solution,
\begin{equation}
    \hh(d_0(t), t) = - s_0(t),
\end{equation}
which states that, at leading-order, the free surface meets the plate at the turnover point \cite{Howison1991}. Finally, matching with the inner region reveals that $\hphi_0 = O(\sqrt{d_0(t) - \hr})$ as $\hr \to d_0(t)^-$, as in the classical Wagner regime \cite{Howison1991}.

Following \cite{Korobkin1982}, it is useful to consider the variational formulation of the axisymmetric problem by introducing the leading-order displacement potential, $\Upsilon_0$, as
\begin{equation}
    \Upsilon_0(\hr, \hz, t) = \hz t + \int_0^t \hphi_0(\hr, \hz, \tau) \dd{\tau},
\end{equation}
which is governed by the equations
\begin{linenomath}
\begin{align}
    \pdv[2]{\Upsilon_0}{\hr} + \frac{1}{\hr} \pdv{\Upsilon_0}{\hr} + \pdv[2]{\Upsilon_0}{\hz} &= 0 &\text{ for } \hz > 0,& \label{eq:bessels_equation}\\
    \pdv{\Upsilon_0}{\hz} &=(t - s_0(t)) - \frac{1}{2} \hr^2 &\text{ on } \hz = 0, \; & \hr < d_0(t), \label{eq:displacement_potential_impermeability} \\
    \pdv{\Upsilon_0}{\hz} &= t + \hh_0(\hr, t) - \frac{1}{2} \hr^2 & \text{ on } \hz = 0, \;& \hr > d_0(t), \label{eq:displacement_potential_kinematic} \\
    \Upsilon_0 & = 0 & \text{ on } \hz = 0, \;& \hr > d_0(t), \label{eq:displacement_potential_dynamic}
\end{align}
\end{linenomath}
such that the displacement potential is $\Upsilon_0 = O((d_0(t) - \hr))^{3/2})$ as $\hr \to d_0(t)^-$. 
The far-field condition for $\hphi_0$ \eqref{eq:outer_far_field} implies that $\Upsilon_0$ is bounded as $\sqrt{\hr^2 + \hz^2} \to \infty$, which means a separable solution for $\Upsilon_0$ can be found via
\begin{equation}
\label{eq:upsilon_solution}
    \Upsilon_0= \int_0^\infty \nu(\lambda, t) e^{- \lambda \hz} J_0(\lambda \hr) \dd{\lambda},
\end{equation}
where $J_0$ is a Bessel function of the first kind and $\nu(\lambda, t)$ is a coefficient function found by solving the dual integral equations necessary to satisfy \eqref{eq:displacement_potential_impermeability} and \eqref{eq:displacement_potential_dynamic}, namely
\begin{equation}
\label{eq:coefficient_function}
    \nu(\lambda, t) = \frac{2}{\pi} \int_0^{d_0(t)}  \sigma \left[\frac{1}{3}\sigma^2 - (t - s_0(t))\right] \sin(\lambda \sigma) \dd{\sigma}.
\end{equation}
We refer to previous studies \cite{Moore2014,Sneddon1966} for further details.

Evaluating \eqref{eq:upsilon_solution} along the plate and expanding as $\hr \to d_0(t)^-$, we find
\begin{equation}
\label{eq:inner_limit_outer_displacement_pot}
\begin{aligned}
    \Upsilon_0(\hr, 0, t) &= \frac{2}{\pi} \sqrt{2 d_0(t)} \left(\frac{1}{3} d_0(t)^2 - (t - s_0(t))\right) \sqrt{d_0(t) - \hr} \\
    &- \frac{8 \sqrt{2} d_0(t)^{3/2}}{9 \pi} (d_0(t) - \hr)^{3/2} + O((d_0(t) - \hr)^{5/2}).
    \end{aligned}
\end{equation}
Hence, to enforce $\Upsilon_0 = O((d_0(t) - \hr)^{3/2})$ as $\hr \to d_0(t)^-$ means that we must have
\begin{equation}
\label{eq:d_solution}
    d_0(t) = \sqrt{3(t - s_0(t))}. 
\end{equation}

Note that if the plate is stationary, $s_0(t) \equiv 0$ and we recover the classical Wagner solution $d_0(t) = \sqrt{3t}$ \cite{Wilson1989}. Clearly, when the plate is compliant, the displacement of the plate is expected to slow down the spreading of the droplet, at least for times early enough that $s_0(t) > 0$. It is well known that the Wagner problem is unstable under time-reversal \cite{Howison1991}, which means the solution breaks down if $\dot{d}_0(t) \leq 0$. We therefore assume that $\dot{s}_0(t) < 1$ in order for the velocity of the turnover point, $\dot{d}_0(t) = \sqrt{3} (1 - \dot{s}_0(t)) / (2 \sqrt{t - s_0(t)})$, to remain positive.

Finally, since it is necessary to calculate the hydrodynamic force on the plate $F(t)$, we use the leading-order form of the Bernoulli equation \eqref{eq:bernoulli} to find the pressure on the plate
\begin{equation}
\label{eq:outer_pressure}
    \hp_0(\hr, 0, t) =  - \pdv[2]{\Upsilon_0}{t} = \frac{4}{9\pi} \dv[2]{}{t} \left[(d_0(t)^2 - \hr^2)^{3/2} \right],
\end{equation}
where $d_0(t)$ is given in terms of $s_0(t)$ in \eqref{eq:d_solution}.

\subsection{Inner region}
Since the pressure is locally singular, there is an inner region moving with the turnover point at $r = \epsilon d(t)$ and the surface of the plate at $z = - \epsilon^2 s(t)$. The appropriate scalings are given by \cite{Howison1991},
\begin{equation}
\begin{gathered}
    r = \epsilon d(t) + \epsilon^3 \tilr, \quad z = - \epsilon^2 s(t) + \epsilon^3 \tz, \\
    \phi = \epsilon^2 \left[ \dot{d}(t) \tilr - \epsilon \dot{s}(t) \tz + \tphi \right] , \quad h = - \epsilon^2 s(t) + \epsilon^3 \tilh, \quad p = \frac{1}{\epsilon^2} \tp. 
\end{gathered}
\end{equation}
Under these scalings, it is straightforward to show that to leading-order, the plate is stationary in the inner region. Hence, as described in detail in \cite{Moore2014}, the leading-order inner problem is quasi-two-dimensional in each plane perpendicular to it, and is given by
\begin{linenomath}
\begin{align}
    \pdv[2]{\tphi_0}{\tilr} + \pdv[2]{\tphi_0}{\tz} &= 0 &\text{for } \tz > \,& 0, \label{eq:inner_laplace}\\
    \pdv{\tphi_0}{\tz} &= 0 &\text{on } \tz =& \, 0, \\
    \pdv{\tphi_0}{\tz} &= \pdv{\tphi_0}{\tilr} \pdv{\tilh_0}{\tilr} &\text{on } \tz =& \, \tilh_0(\tilr, t), \\
    \left(\pdv{\tphi_0}{\tilr}\right)^2 + \left(\pdv{\tphi_0}{\tz}\right)^2 &= \dot{d}_0(t)^2 &\text{on } \tz =& \, \tilh_0(\tilr, t) \label{eq:inner_dynamic},
\end{align}
\end{linenomath}
subject to appropriate matching conditions into the outer region,
\begin{equation}
    \tphi_0 \sim - \dot{d}_0(t) \tilr + O(\sqrt{\tilr^2 + \tz^2}) \text{ as } \sqrt{\hr^2 + \hz^2} \to \infty,
\end{equation}
and towards the splash sheet
\begin{equation}
    \tilh_0(\tilr, t) \to J(t) \text{ as } \hr \to \infty,
\end{equation}
where $J(t)$ is referred to as the asymptotic sheet thickness. 

The solution to this problem is well known \cite{Wagner1932}, and given parametrically by
\begin{equation}
\label{eq:inner_solution}
\begin{aligned}
    \tphi_0 = a(t) - \frac{\dot{d}_0(t) J(t)}{\pi} (1 + \mathfrak{R}[\zeta + \log(\zeta)]), \\
    \tilr + i \tz = \frac{J(t)}{\pi} \left[\zeta + 4 i \sqrt{\zeta} - \log(\zeta) + i \pi -1 \right],
\end{aligned}
\end{equation}
where $\zeta \in \mathbb{C}$, $\mathfrak{I}(\zeta) > 0$, $a(t) \in \mathbb{C}$ is an arbitrary function of time, the branch cuts for $\log(\zeta)$ and $\sqrt{\zeta}$ are taken along $\mathfrak{R}(\zeta) < 0, \mathfrak{I}(\zeta) = 0^+$ and $\mathfrak{R}$, $\mathfrak{I}$ denote the real and imaginary parts of a complex number. 

To match with the leading-order outer solution, we take $|\zeta| \to \infty$ in \eqref{eq:inner_solution}, which yields
\begin{equation}
\label{eq:outer_limit_of_inner}
    \tphi_0 \sim - \dot{d}_0(t) \tilr + 4 \dot{d}_0(t) \frac{J(t)}{\pi} \mathfrak{R} \left[i \sqrt{\tilr + i \tz}\right].
\end{equation}
Thus, comparing \eqref{eq:outer_limit_of_inner} with \eqref{eq:inner_limit_outer_displacement_pot} gives the leading-order sheet thickness
\begin{equation}
\label{eq:J_solution}
    J(t) = \frac{2 d_0(t)^3}{9 \pi} = \frac{2}{\sqrt{3} \pi} (t - s_0(t))^{3/2}. 
\end{equation}
Again note that as $d_0(t) = \sqrt{3(t - s_0(t))}$, the displacement of the plate slows the spreading of the droplet, which leads to a thinner splash sheet. This is consistent with the findings of \cite{Howland2016a}, who showed that soft substrates inhibit splashing. Note that the derivative of the sheet thickness $\dot{J}(t) = \sqrt{3} (1 - \dot{s}_0(t))\sqrt{t - s_0(t)} / \pi$ is positive for all $t$, so the sheet thickness will still increase for all time within the Wagner model.

The leading-order pressure in the inner region is 
\begin{equation}
    \tp_0 = - \frac{1}{2} \left[\left(\pdv{\tphi_0}{\tilr}\right)^2 + \left(\pdv{\tphi_0}{\tz} \right)^2 - \dot{d}_0(t)^2 \right].
\end{equation}
Along the surface of the plate, where $\tz = 0$, this solution is given parametrically by 
\begin{equation}
\label{eq:inner_pressure}
    \tp_0(\tilr, 0, t) = \frac{2 \dot{d}_0(t)^2 e^{\eta}}{(1 + e^{\eta})^2}, \; \tilr = \frac{-J(t)}{\pi} \left[e^{2\eta} + 4 e^{\eta} + 2 \eta + 1 \right] \; \mbox{for} \; -\infty<\eta<\infty,
\end{equation}
where $d_0(t)$ and $J(t)$ are given in terms of $s_0(t)$ in \eqref{eq:d_solution} and \eqref{eq:J_solution} respectively. 

\subsection{Splash sheet region}
\label{subsec:splash_sheet}
Upon impact, the fluid is ejected from the inner region into a thin, fast-moving sheet of fluid attached to the plate. In this region, we rescale
\begin{equation}
    r = \epsilon \br, \; z = - \epsilon^2 s(t) + \epsilon^3 \bz, \; h = - \epsilon^2 s(t) + \epsilon^3 \bh, \; \phi = - \epsilon^2 \dot{s}(t) \bz + \bphi, \; p = \epsilon \bp.
\end{equation}
As described in detail by \cite{Oliver2002}, the leading-order splash sheet problem for the radial velocity $\bu_0 = \partial \bphi_0 / \partial \br$ and free surface height $\bh_0$ reduces to the zero-gravity shallow-water equations. These equations can be solved using the method of characteristics and the solution is, as derived in \cite{Moore2014,Oliver2002}, 
\begin{equation}
\label{eq:splash_sheet_solution}
    \br = 2 \dot{d}_0(\tau) (t - \tau) + d_0(\tau), \;\bu_0 = 2 \dot{d}_0(\tau), \; \bh_0 = \frac{\dot{d}_0(\tau) J(\tau)}{\dot{d}_0(\tau) - 2 \ddot{d}_0(\tau) (t - \tau)},
\end{equation}
where $0 < \tau < t$. 

The subsequent solution for the pressure in the splash sheet region is found by differentiating the Bernoulli equation \eqref{eq:bernoulli} with respect to $z$, expressing in the splash sheet variables, and expanding to leading-order, such that
\begin{equation}
    \pdv{\bp_0}{\bz} = \ddot{s}_0(t) - \dot{s}_0(t) \pdv{\bu_0}{\br}.
\end{equation}
Integrating with respect to $\bz$ and noting that $\bp_0 = 0$ at $\bz = \bh_0$ means the leading-order pressure along the plate is
\begin{equation}
\label{eq:splash_sheet_pressure}
    \bp_0(\br, 0, t) = \left(\dot{s}_0(t) \pdv{\bu_0}{\br} - \ddot{s}_0(t)\right) \bh_0(\br, t).
\end{equation}
It is worth noting that in the classical case of a stationary plate, where $s_0(t) \equiv 0$, the leading-order pressure \eqref{eq:splash_sheet_pressure} would be zero and instead $\bp = O(\epsilon^2)$. Therefore the velocity $\dot{s}_0(t)$ and acceleration $\ddot{s}_0(t)$ of the plate increase the magnitude of the pressure in the splash sheet region. In particular, if $\dot{s}_0(t) \partial \bu_0 / \partial \br - \ddot{s}_0(t) < 0$, the leading-order pressure would be below atmospheric pressure, which could provide a possible mechanism for splash sheet detachment. However, the contribution the pressure in the splash sheet region makes to the leading-order force is still $O(\epsilon^3)$, which is lower in magnitude than the contributions from the outer and inner regions, so we shall neglect it henceforth in this analysis.

\subsection{Composite pressure}
\label{subsec:composite_pressure}

Classically, the hydrodynamic force is determined by integrating only the outer pressure \eqref{eq:outer_pressure} for $0 \leq \hr < d_0(t)$ \cite{Oliver2002}. However, as will be discussed in \textsection \ref{sec:conclusion}, we find better agreement with the numerical simulations at later times by using the composite expansion between the outer and inner regions. Van Dyke's matching principle \cite{VanDyke1975} is used to find the overlap function between the outer and inner solutions for $r < \epsilon d_0(t)$. By evaluating the time derivatives in \eqref{eq:outer_pressure}, the one-term-outer pressure is
\begin{equation}
    \text{(1.t.o)}p(r, 0, t) = \frac{1}{\epsilon} \left[-\frac{4(\hr^2 - 2 d_0(t)^2) \dot{d}_0(t)^2}{3 \pi \sqrt{d_0(t)^2 - \hr^2}} + \frac{4 d_0(t) \ddot{d}_0(t)}{3 \pi} \sqrt{d_0(t)^2 - \hr^2} \right].
\end{equation}
Expressing this in the inner variables and expanding to leading order gives the overlap pressure
\begin{equation}
\label{eq:overlap_pressure}
\begin{aligned}
    p_\text{overlap}(r, 0, t) = \text{(1.t.i)(1.t.o)}p(r, 0, t) 
    = \frac{2 \sqrt{2} d_0(t)^{3/2} \dot{d}_0(t)^2}{3 \pi \epsilon^2 \sqrt{- \tilr}}
    = \frac{2 \sqrt{2} d_0(t)^{3/2} \dot{d}_0(t)^2}{3 \pi \sqrt{\epsilon} \sqrt{\epsilon d_0(t) - r }}.
\end{aligned}
\end{equation}
Therefore the composite expansion for $p(r, 0, t)$ is
\begin{equation}
\begin{aligned}
\label{eq:composite_pressure}
    p_{\text{comp}}&(r, t) = H(\epsilon d_0(t) - r) \left[\frac{1}{\epsilon} \hp_0(r / \epsilon, 0, t) - p_\text{overlap}(r, 0, t) \right]   \\
    &+ \frac{1}{\epsilon^2} \tp_0(r / \epsilon^3 - d_0(t)/ \epsilon^2, 0, t),
\end{aligned}
\end{equation}
where $H$ is the Heaviside step function, $\hp_0$ is given by \eqref{eq:outer_pressure} and $\tp_0$ by \eqref{eq:inner_pressure}. 

\begin{figure}[t]
    \centering
    \includegraphics[width=\textwidth]{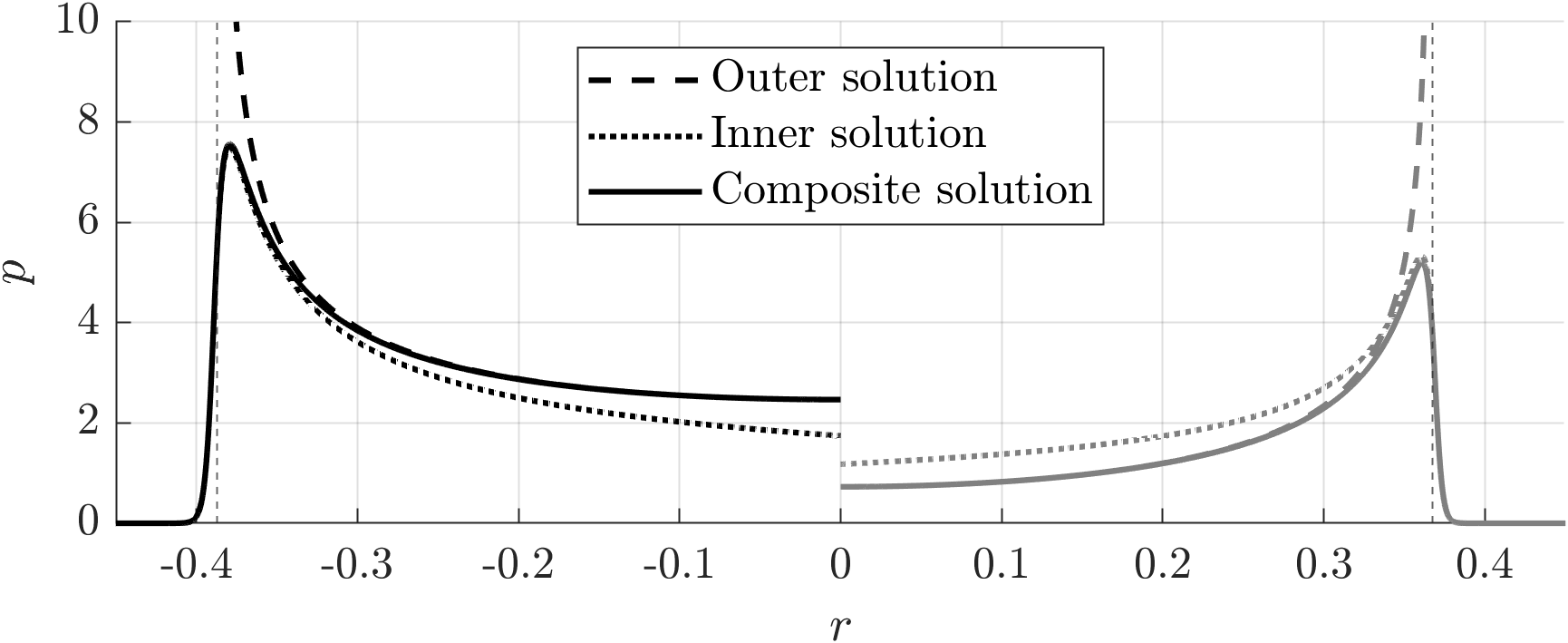}
    \caption{Analytical solutions for the pressure for $\epsilon = 0.1$ at $t = 5$, with the left-hand side black lines showing the stationary plate case, $s_0(t) = 0$ (reflected in $r$), and the right-hand side grey lines the moving plate case, $s_0(t) = 0.02 t^2$. The outer solution for the pressure \eqref{eq:outer_pressure} is shown by the dashed lines, the inner solution \eqref{eq:inner_pressure} by the dotted lines and the composite solution \eqref{eq:composite_pressure} by the solid lines. The thin vertical dashed lines show the location of the turnover point $r = \epsilon d_0(t)$. }
    \label{fig:analytical_pressure_comparison}
\end{figure}

The composite pressure profile \eqref{eq:composite_pressure} depends on the plate displacement $s_0(t)$, which is solved for in \textsection \ref{sec:plate_displacement_solution} once the hydrodynamic force is determined. However, in order to illustrate the effects of a moving substrate on the pressure, we compare its value in the case where the plate is stationary ($s_0(t) \equiv 0$) to that of a prescribed moving plate case where $s_0(t) = 0.02 t^2$. Note that this value for $s_0(t)$ is chosen for illustrative purposes and is not a solution to \eqref{eq:dimensionless_plate_ode}, and only satisfies the assumption that $\dot{s}_0(t) < 1$ for $t < 25$. 

We compare the solutions at $t=5$ in Fig. \ref{fig:analytical_pressure_comparison}, where the left-hand side black lines  shows the outer, inner and composite solutions to the pressure for the stationary plate case (reflected in $r$), and the right-hand side grey lines shown the corresponding values for the moving plate case. It can be seen that the turnover point in the moving plate case has advanced less than in the stationary plate case, as according to \eqref{eq:d_solution}. The pressure in the moving plate case is significantly lower overall than in the stationary plate case. This shows how the downwards motion of the plate does not just slow the spreading, but also decreases the hydrodynamic pressure inside the droplet. Also noteworthy is that the inner solution under-estimates the solution away from the turnover point in the stationary plate case, but over-estimates it in the moving plate case. 

\subsection{Hydrodynamic force}
\label{subsec:hydrodynamic_force}
In order to solve \eqref{eq:dimensionless_plate_ode} for the displacement of the plate $s(t)$, the value of the hydrodynamic force, $F(t)$, needs to be determined to leading order. We approximate the force by integrating the composite expansion to the pressure \eqref{eq:composite_pressure} across the outer and inner regions. The composite expansion for the force is hence 
\begin{equation}
\label{eq:composite_force}
\begin{aligned}
    F_\text{comp}(t) &= \frac{8}{9} \epsilon d_0(t)^3 ((4- 2 \sqrt{2})\dot{d}_0(t)^2 + \ddot{d}_0(t)d_0(t)) \\
    &+ \frac{8 \epsilon^4 \dot{d}_0(t)^2 J(t)^2}{\pi}  e^{\eta_0(t)}\left[\frac{\pi d_0(t)}{\epsilon^2 J(t)} +1 - \frac{1}{3} e^{2 \eta_0(t)} - 2 e^{\eta_0(t)} - 2 \eta_0(t)\right],
\end{aligned}
\end{equation}
where $\eta_0(t)$ is defined implicitly by
\begin{equation}
\label{eq:eta_0_def}
    e^{2 \eta_0(t)} + 4 e^{\eta_0(t)} + 2 \eta_0(t) + 1 = \frac{\pi d_0(t)}{\epsilon^2 J(t)},
\end{equation}
where the detailed derivation can be found in Appendix \ref{appendix:composite_force}. 

It is worth noting that the composite force $F_\text{comp}(t)$ differs from the force on a stationary plate if and only if the turnover point $d_0(t)$ or sheet thickness $J(t)$ differ from their corresponding stationary plate values.

\subsection{Plate displacement solution}
\label{sec:plate_displacement_solution}
The remaining unknown from \textsection \ref{subsec:outer_region} - \ref{subsec:hydrodynamic_force} is the leading-order plate displacement $s_0(t)$, where $s_0(t)$ appears in the solution \eqref{eq:d_solution} for the turnover point $d_0(t)$  and in the solution \eqref{eq:J_solution} for the jet-thickness $J(t)$. The plate displacement is found by solving the second order ordinary differential equation \eqref{eq:dimensionless_plate_ode}, approximating the force term $F(t)$ by the composite force $F_\text{comp}(t)$ \eqref{eq:composite_force}. The resulting equation is non-linear and implicit, and is solved using MATLAB's ode15i solver in conjunction with the fsolve solver to find $\eta_0(t)$ via \eqref{eq:eta_0_def} at each timestep. As the value of $\dot{d}_0(t)$ diverges at $t = 0$, the numerical scheme is initialised at a time $t = t_i = 10^{-9}$, with zero initial guesses for $s_0(t_i) = \dot{s}_0(t_i) = \ddot{s}_0(t_i) = 0$. A small-time asymptotic analysis of the plate displacement reveals that $s_0(t) = O(t^{5/2})$ as $t \to 0$, so the problem is regular and we are hence justified in taking zero initial guesses. The results will be discussed in comparison to the full DNS in \textsection \ref{sec:results}.

For $\epsilon = 0.1$, the numerical solution for $s_0(t)$ is found for $0 \leq t \leq 100$ on 1 CPU in approximately 10 seconds. In comparison, the DNS results in \textsection \ref{sec:results} required approximately $24$ CPU hours for the same dimensionless timescale, hence finding a numerical approximation to the analytical solution is significantly less computationally expensive than the DNS and a valuable first incursion into the parameter space, providing much-needed direction for the heavier numerical machinery. 

\section{Direct numerical simulations}
\label{sec:dns}
We build on the open-source, volume-of-fluid package \texttt{Basilisk} \cite{Popinet2015} to implement this complex multi-phase system, retaining effects due to viscosity and density in both the liquid and the gas, as well as surface tension and gravity. \texttt{Basilisk}, and its predecessor \texttt{Gerris} \cite{Popinet2003}, have been used extensively to study interfacial flows, and in particular droplet impact, with great success over the past two decades, cross-fertilising investigative efforts within experimental, analytical and computational communities alike \cite{Cimpeanu2018,Lopez-Herrera2019,Philippi2016,Thoraval2012,Wildeman2016}.

\subsection{Moving frame coordinates}
\label{subsec:moving_frame_coordinates}
In order to avoid including an embedded boundary in the quadtree-structured computational domain, we transfer the flow into a frame moving with the plate, fixing the plate along the bottom computational boundary. The dimensionless moving frame coordinates are defined by $\vb{x}' = (x', y', z') = \vb{x} + \vb{s}(t)$, $\vb{u}' = \vb{u} + \dot{\vb{s}}(t)$, where $\vb{s}(t) = (0, 0, s(t))$, and the prime $'$ decorates all quantities in the moving frame. 
Introducing the dimensionless variable density $\rho'(\vb{x}', t)$ and viscosity $\mu'(\vb{x}', t)$, such that, following notation from \textsection \ref{subsec:non_dimensionalisation}, $\rho' = 1$, $\mu' = 1$ in the liquid and $\rho' = \rho_R$, $\mu' = \mu_R$ in the gas, the dimensionless governing equations in the moving frame are given by
\begin{linenomath}
\begin{align}
    \rho' \left( \pdv{\vb{u}'}{t} + (\vb{u}' \vdot \grad') \vb{u}' \right) &= - \grad' p' + \frac{\mu'}{\text{Re}} (\grad')^2 \vb{u}' + \frac{\kappa' \delta_s'}{\text{We}}  \vu{n}' + \rho' \ddot{\vb{s}}(t) - \frac{\rho'}{\text{Fr}^2} \vu{n}_z', \label{eq:moving_momentum}\\
    \grad' \vdot \vb{u}' &= 0, \label{eq:moving_continuity}\\
    u_{z'}' &= 0 \text{ for } z' = 0, \; x'^2 + y'^2 < R_p^2, \label{eq:moving_kinematic}
\end{align}
\end{linenomath}
where $\delta_s'$ is a Dirac distribution centred on the liquid-gas free surface and $\vu{n}'$ is the unit normal to the interface. Note that the kinematic condition \eqref{eq:moving_kinematic} is that of a stationary plate, and the problem in the moving frame is equivalent to a droplet impacting onto a stationary plate, with the liquid and gas under additional forcing equal to $\rho' \ddot{\vb{s}}(t)$. In the far-field, we assume the pressure tends to 0 (neglecting variations due to gravity) and the vertical velocity tends to $\vb{u}' \to \dot{\vb{s}}(t)$. The prime notation is dropped for the remainder of this section for brevity.

\subsection{Computational setup}
\label{subsec:computational_setup}
In all our simulations, we consider a droplet with dimensional radius $R^*_d = 1$ mm initially travelling vertically downwards at speed $V^* = 5$ m/s, where the values of Re, We and Fr$^2$, $\rho_R$, $\mu_R$, are given in \textsection \ref{subsec:assumptions}. The radius of the plate is taken to be twice the initial radius of the droplet, so that $R_p = 2$, and the initial separation between the bottom of the droplet and top of the plate is $\delta^* = 0.125 R^*_d = 0.125$ mm. The values of the mass ratio $\alpha$, damping factor $\beta$ and stiffness factor $\gamma$ are varied across different parametric studies.

\begin{figure}
    \centering
    \includegraphics[width=1.0\textwidth]{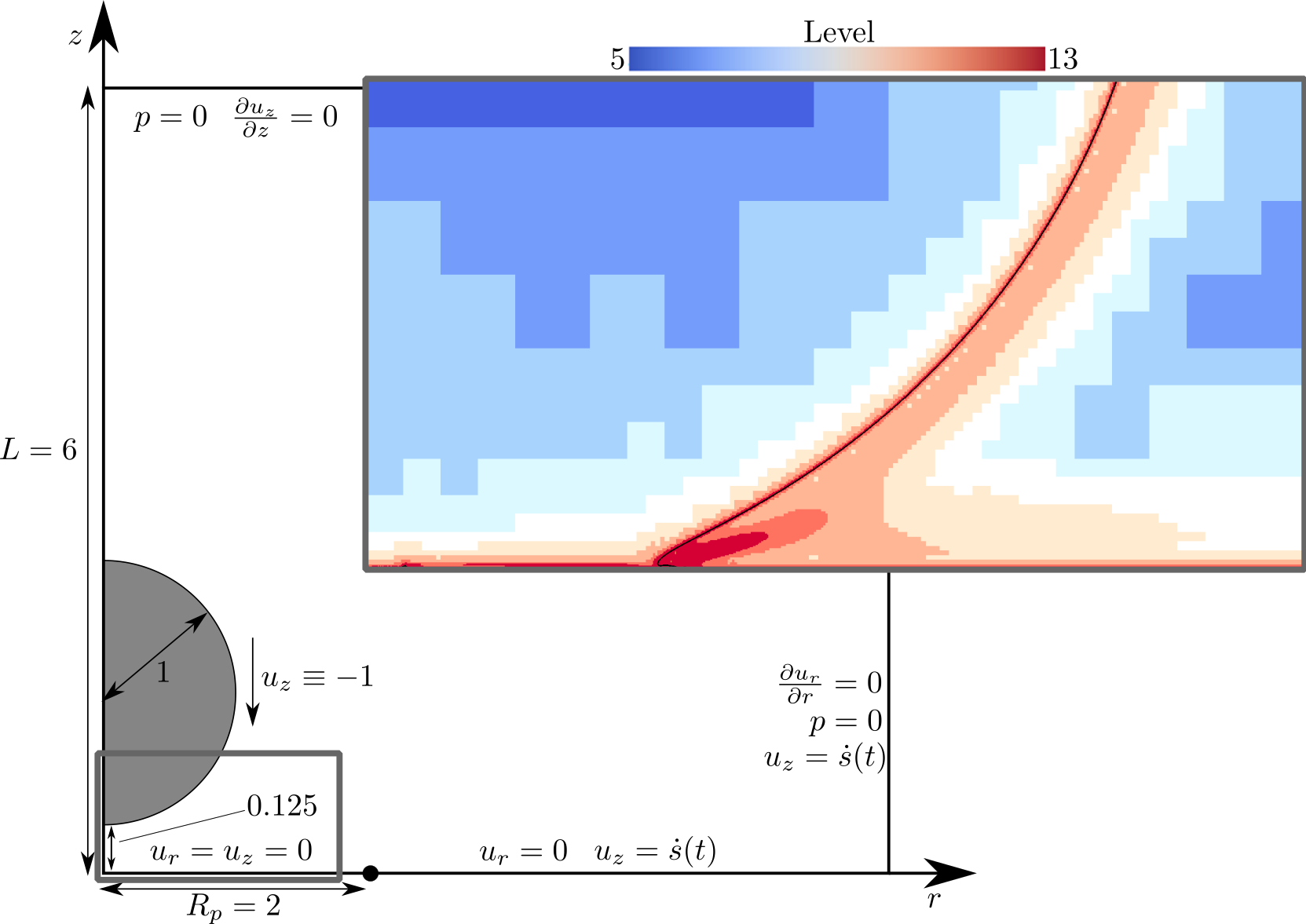}
    \caption{Direct numerical simulation setup at $t = t_0 = -0.125$, for a droplet of unit non-dimensional radius, separated from the plate by a distance of $0.125$ and travelling with a vertical velocity $-1$. The inset shows a snapshot of a simulation at $t = 0.045$ close to the surface, around the area indicated with the grey rectangle. The colour map illustrates the adaptive mesh refinement strategy, while the black line depicts the location of the interface.}
    \label{fig:dns_setup}
\end{figure}

The dimensionless governing equations \eqref{eq:moving_momentum}-\eqref{eq:moving_continuity} are solved using \texttt{Basilisk} within an axisymmetric domain, where the droplet initially has unit dimensionless radius, travelling with uniform vertical velocity $-1$. The axisymmetric computational domain for the simulations is shown in Fig. \ref{fig:dns_setup}. The domain is given by a square box, with the $r$ axis along the bottom boundary and the $z$ axis along the left boundary. The side length of the domain is set to $L = 6$, which is sufficiently large so that far-field conditions do not artificially alter the target dynamics. Neumann conditions $\partial_n \vb{u} = 0$ are specified along the top and right boundaries, where $\partial_n$ is the partial derivative in the normal direction to the boundary. The vertical velocity along the right boundary is specified as $u_z = \dot{s}(t)$, to reflect the far-field velocity condition. The appropriate symmetry conditions are applied along the left boundary. A mixed boundary condition is specified along $z = 0$, with $u_z = 0$ for $r \leq R_p$ and $u_z = \dot{s}(t)$ for $r > R_p$, the former representing the kinematic condition along the plate \eqref{eq:moving_kinematic} and the latter representing the far-field condition. The no-slip condition $u_r = 0$ and a 90$^{\circ}$ contact angle are applied along the bottom boundary. Finally, the pressure $p$ in the gas is set to to be zero along the top and right boundaries in line with the far-field condition \eqref{eq:gas_far_field}. 

The quadtree grid construction features in \texttt{Basilisk} allow for a high grid resolution in areas of interest, varying from levels 5 to 13, where level $n$ corresponds to $2^n$ square cells per dimension, if the grid were uniform. In this case, the largest cell has side length $6 / 2^5$, corresponding to $0.188$ mm in dimensional terms. The smallest cell has side length $6 / 2^{13}$, corresponding to $0.732\ \mu$m. In order to accurately calculate the force on the plate, a region along the bottom boundary for $0 \leq r \leq R_p$ of height $24/2^{13}$ ($\approx 2.93 \ \mu $m in dimensional terms) is held at level 13, such that the bottom four grid cells in this region are at maximum level. 
This avoids numerical errors induced by multi-grid interpolation in a region which requires particular care due to the delicate fluid-structure interaction calculations outlined below. 
Adaptive mesh refinement is also used to refine the domain in regions where the velocities and interface location are rapidly changing. An example of the typical grid structure is shown in the inset of Fig. \ref{fig:dns_setup}.

At regular timesteps $\Delta t = 10^{-4}$ (corresponding to $20$ ns in dimensional terms), the hydrodynamic force applied to the surface of the plate, $F(t)$, is determined by numerically integrating the pressure, $p$, and the viscous stress, $-2 \mu \partial u_z / \partial z$, along the bottom boundary for $0 \leq r \leq R_p$. This value of the force is then used to solve the dimensionless plate displacement equation \eqref{eq:dimensionless_plate_ode} using a second-order finite difference scheme, giving $s(t), \dot{s}(t)$ and $\ddot{s}(t)$. The boundary conditions are then updated with the new value of $\dot{s}(t)$, and the vertical acceleration in all of the cells is incremented by $\rho \ddot{s}(t)$. 

Several computational details are noteworthy in terms of ensuring a robust fluid-structure interaction calculation procedure. As observed in other studies on droplet impact \cite{Philippi2016}, numerical instabilities in the projection solver used for the resulting Poisson equation within \texttt{Basilisk} may cause the calculated pressure values to fluctuate between timesteps, thus causing the resulting force values to vary artificially. These pressure spikes lead to artifacts in the finite difference scheme, which can ultimately result in the simulation breaking down due to diverging acceleration terms. To prevent this, we use a peak-detection algorithm \cite{VanBrakel} to identify numerical spikes and smooth out the resulting force. Spatial filtering is also used to manage the contrast in density and viscosity between the liquid and gas phases. Furthermore, any small gas bubbles or liquid drops that have a diameter smaller than sixteen level $13$ cells (corresponding to $\approx 10\ \mu$m in dimensional terms) are deemed unphysical and dynamically removed, with the exception of the entrapped gas bubble centred at $r = z = 0$. 

The simulations span $0.8$ dimensionless time units, corresponding to $0.2$ ms in dimensional time. During this timescale, the end of the splash sheet typically reaches $r \approx 1.9$, close to the edge of the plate, and the turnover point reaches $r \approx 1.3$. The early impact stage can be considered over long before the turnover point surpasses the initial droplet radius, hence we also capture timescales beyond when we expect the analytical results to be valid. Each individual simulation consisted in approximately $60,000$ (dynamically adapted) degrees of freedom and was executed in parallel on $4-8$ CPUs, for approximately $24$ CPU hours on local high performance computing facilities.

\subsection{Numerical validation}
As will be demonstrated in the following section, the excellent agreement between analytical and numerical results gives us encouragement that the simulations are converging to the correct solution. However to ensure computational robustness we have also conducted a comprehensive validation study, confirming that the results in \textsection \ref{sec:results} are mesh-independent at the current level, as well as insensitive to further increases in the computational domain size. For all validation tests, the mass-ratio was chosen to be $\alpha = 100$, with the damping and stiffness factor $\beta = \gamma = 0$. For mesh-independence, we ran multiple simulations varying the maximum refinement level from 9 to 13, finding that the calculated force did not vary noticeably beyond level 12. In dimensional terms, this means the smallest cells must have a side length of at most $\approx$ 1.46 $\mu$m. Similarly, the width of the computational domain was varied for $L = $ 3, 6, 12 and 24. We found that after $L = 6$, the calculated force value no longer changed to a significant degree. We are thus confident in proceeding with a comprehensive parametric study, exploring the solution space with both analytical and computational approaches.

\section{Results and comparisons}
\label{sec:results}
The aim of this section is two-fold: firstly, to systematically compare the predictions of the analytical model from \textsection \ref{sec:analytical} to the results of the numerical simulations from \textsection \ref{sec:dns}, identifying timescales during which good agreement is observed and, secondly, to provide insight into the physical mechanisms introduced once substrate motion is allowed, systematically showing how the mass ratio $\alpha$, damping factor $\beta$ and stiffness factor $\gamma$ affect the dynamics of the system. To facilitate the comparison of the analytical and numerical results, we re-express all quantities into the original non-dimensional variables from \textsection \ref{subsec:non_dimensionalisation}, transforming from the asymptotic variables $t = \epsilon^2 \hat{t}$ in \textsection \ref{subsec:asymptotic_governing} and the primed moving frame variables in \textsection \ref{subsec:moving_frame_coordinates}. All simulations were conducted for $-0.125 \leq t \leq 0.675$.

\subsection{Stationary plate comparison}
\label{subsec:stationary_plate_comparison}
\begin{figure}
    \centering
    \begin{subfigure}[t]{1.0\textwidth}
        \includegraphics[width=\textwidth]{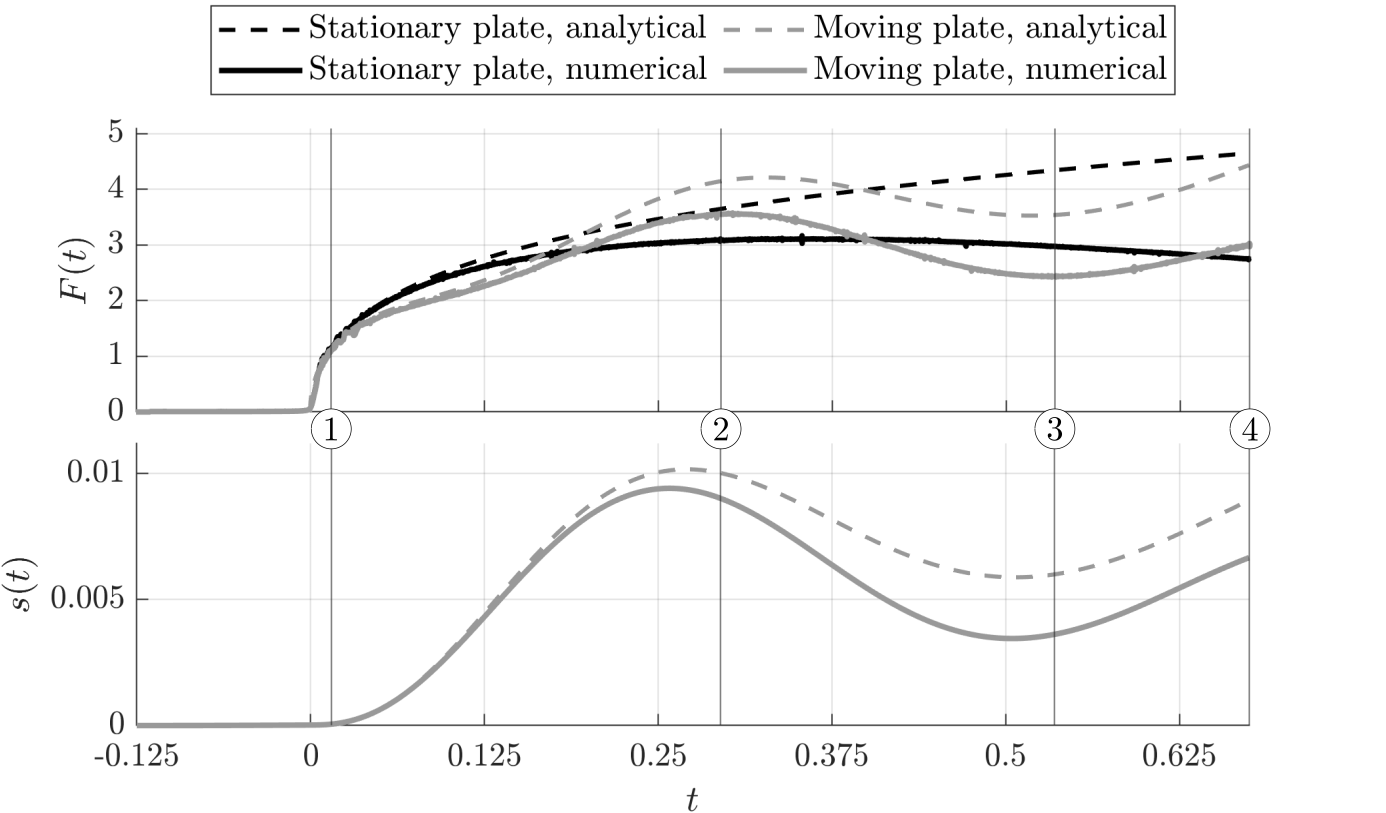}
    \end{subfigure}
    \begin{subfigure}[t]{0.48\textwidth}
        \includegraphics[width=\textwidth]{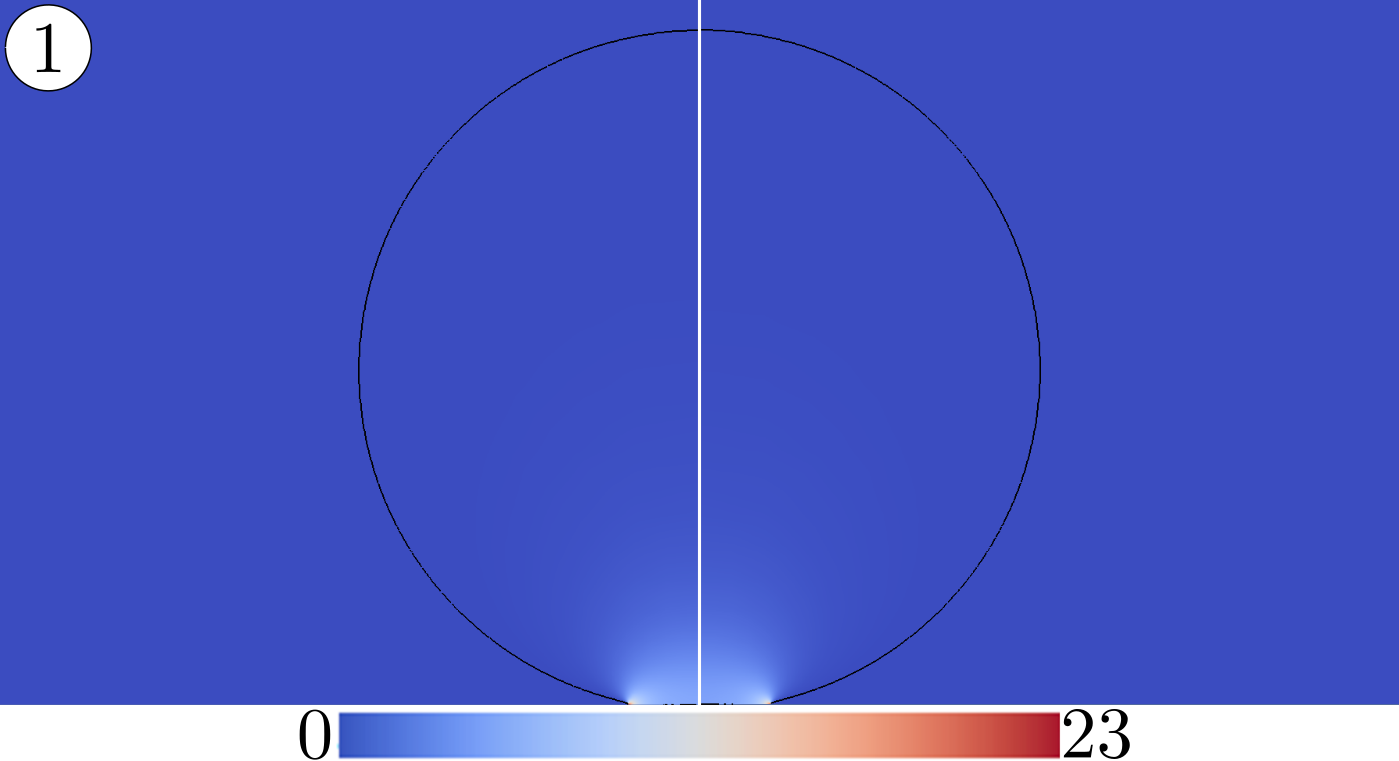}
        \caption{$t = 0.015$}
    \end{subfigure}
    \begin{subfigure}[t]{0.48\textwidth}
        \includegraphics[width=\textwidth]{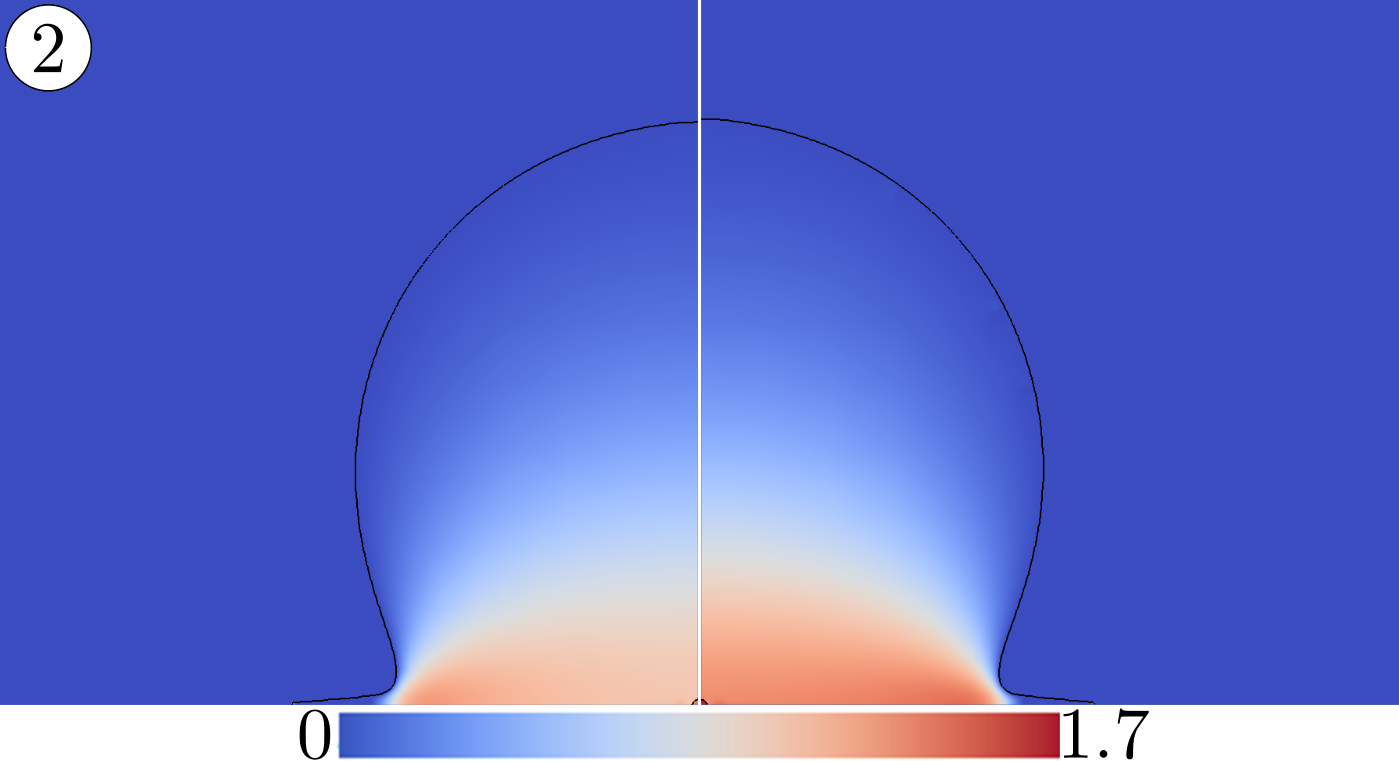}
        \caption{$t = 0.295$}
    \end{subfigure}
    \begin{subfigure}[t]{0.48\textwidth}
        \includegraphics[width=\textwidth]{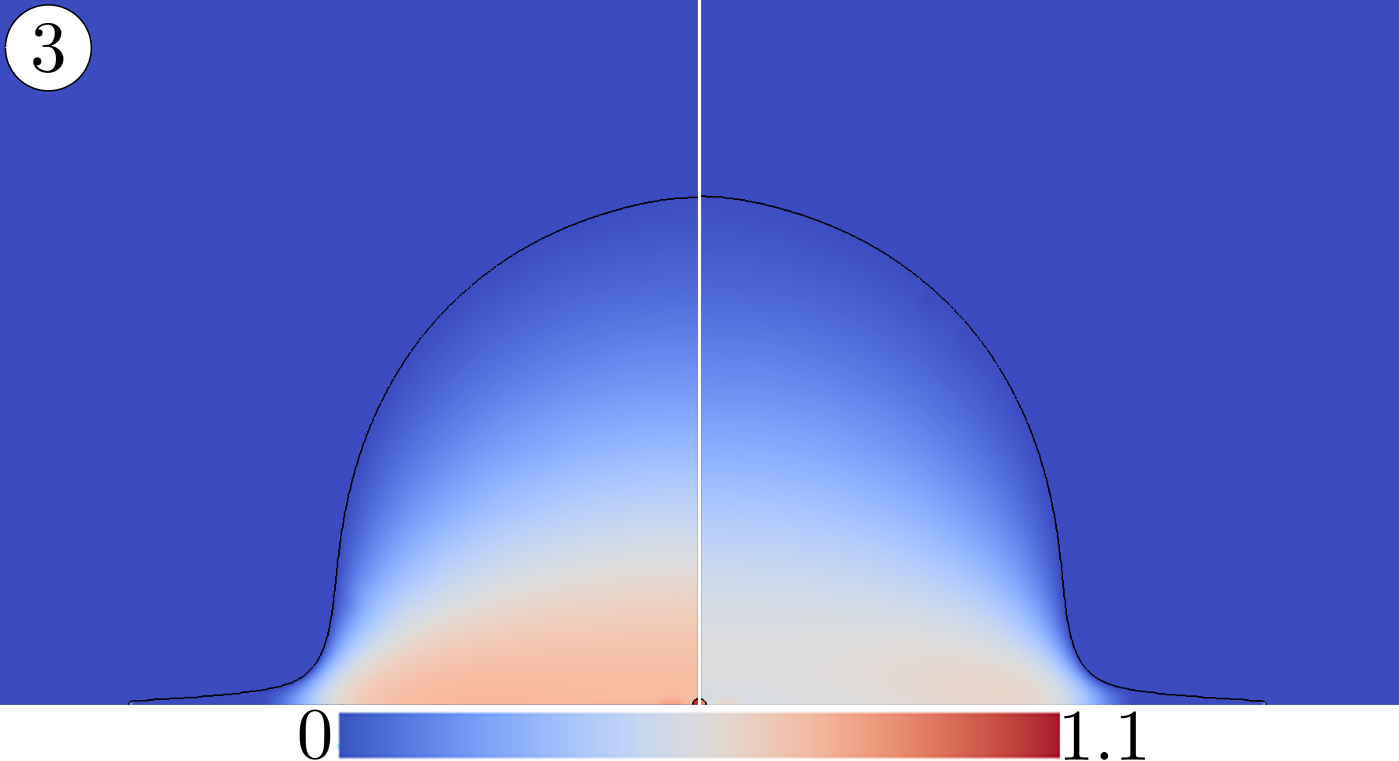}
        \caption{$t = 0.535$}
    \end{subfigure}
    \begin{subfigure}[t]{0.48\textwidth}
        \includegraphics[width=\textwidth]{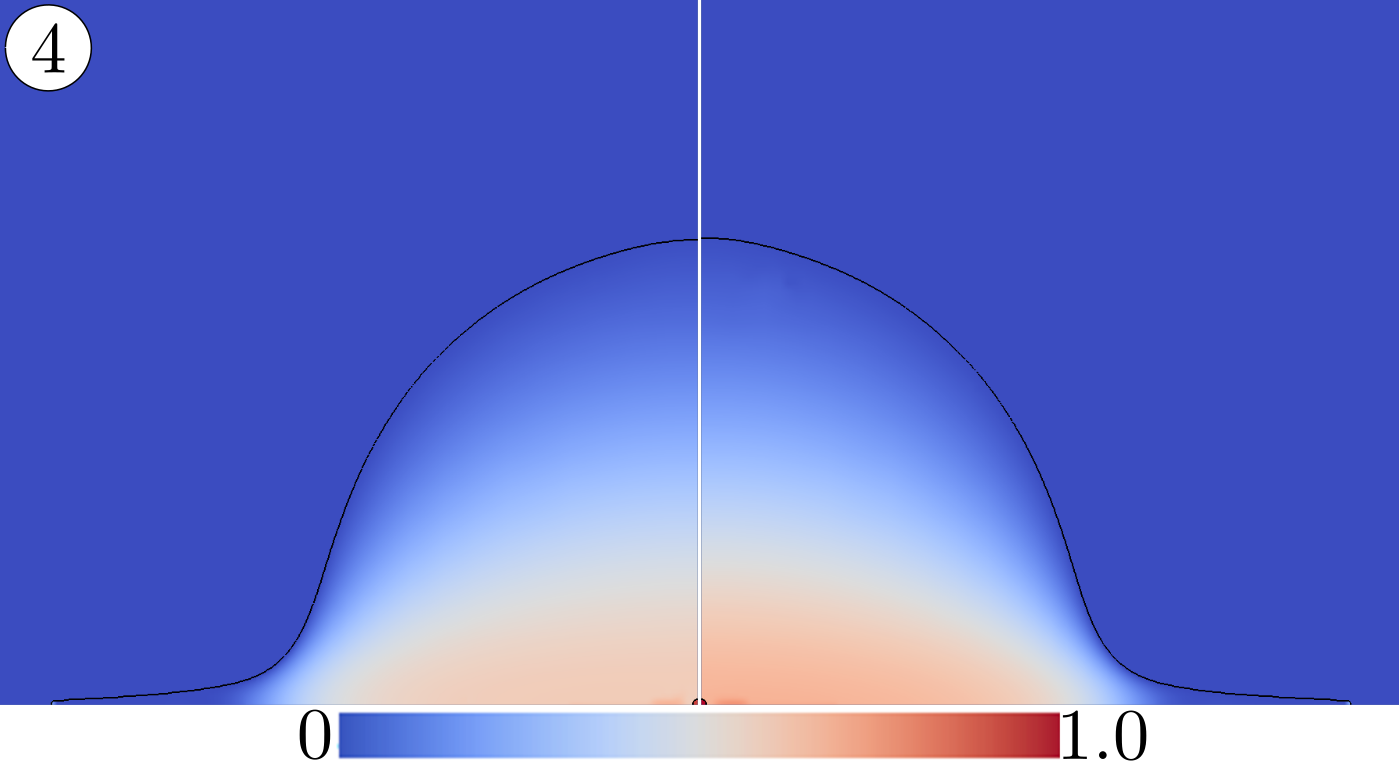}
        \caption{$t = 0.675$}
    \end{subfigure}
    \caption{(Top) Comparison of the hydrodynamic force $F(t)$ between the stationary plate case (black) and a moving plate (grey) with mass ratio $\alpha = 2$, damping factor $\beta = 0$, stiffness factor $\gamma = 500$, with a dashed line for the analytical solution from \eqref{eq:composite_force} and a solid line for the corresponding numerical value. (Middle) Displacement of the moving plate case $s(t)$, with the dashed line showing the analytical solution to \eqref{eq:dimensionless_plate_ode} and the solid line depicting the corresponding numerical results. (Lower panels) Comparison between the pressure $p$ of the DNS between the stationary plate case (left) and the moving plate case (right) at the times labelled 1-4 in the plots above.}
    \label{fig:force_and_displacement_comparison}
\end{figure}

In order to understand the influence the motion of the plate has on the system, we must compare to the case where the plate is held in a stationary position. In particular, we wish to find where the hydrodynamic force on the plate $F(t)$ differs from the corresponding value for the stationary plate case, allowing us to identify where the plate motion has a strong influence on the dynamics of the droplet. 

The analytical and numerical predictions for the hydrodynamic force $F(t)$ and the plate displacement $s(t)$ for $\alpha = 2$, $\beta = 0$, $\gamma = 500$ are shown in Fig. \ref{fig:force_and_displacement_comparison}, alongside the corresponding analytical and numerical predictions for the stationary plate case. Under no damping or hydrodynamic forcing, the plate displacement $s(t)$ in \eqref{eq:dimensionless_plate_ode} would oscillate about $s(t) = 0$ with a natural time period $T = 2 \pi \sqrt{\alpha / \gamma} \approx 0.397$. The parameters $\alpha, \gamma$ are chosen so $T$ is of the same order of magnitude as the timescale of the simulation $-0.125 \leq t \leq 0.675$. In dimensional terms, this system corresponds to an aluminium plate of radius 2 mm, thickness $\approx$ 0.06 mm and spring constant $k^* \approx$ 12.5 N/m. In Fig. \ref{fig:force_and_displacement_comparison}, snapshots of the simulations at the points in time labelled 1-4 in the graphs are shown in the panels, with the left-hand panels showing the stationary plate case, the right-hand panels showing the moving plate case and the colour map showing the pressure distribution in each. The computed value of the viscous stress along the plate in the DNS was typically $< 0.1 \%$ of the pressure, hence the dominant contribution to the numerical results for the force $F(t)$ was due to the pressure itself. 

Point 1 ($t = 0.015$) is shortly after the impact of the droplet. Here, the value of $F(t)$ for the moving plate is close to that of the stationary plate, as the plate has only deformed to within a distance of $O(10^{-3})$, and it can be seen in the panels in Fig. \ref{fig:force_and_displacement_comparison} that the pressure distribution for both is similar. The plate displaces downwards until around $t = 0.25$, at which point the strength of the elastic restoring force, $\gamma s(t)$, causes the plate to move back upwards. Shortly after this, at point 2 ($t = 0.295$), the graphs in Fig. \ref{fig:force_and_displacement_comparison} show that the hydrodynamic force in the moving plate case is greater than the stationary plate case, due to the stronger pressure that can be seen in the panels. The plate subsequently moves upwards into the droplet until around $t =0.5$, when the elastic force balances with the hydrodynamic force and the plate begins to accelerate downwards again. The hydrodynamic force reaches a local minimum at point 3 ($t = 0.535$). Point 4 ($t = 0.675$) marks the end point of the simulations, but it can be extrapolated from the graphs in Fig. \ref{fig:force_and_displacement_comparison} that this oscillatory behaviour would continue for later times.

The analytical solutions in Fig. \ref{fig:force_and_displacement_comparison} show excellent agreement with the numerical results up until close to point 2 at $t = 0.295$. This is remarkable, as the analytical model makes the assumption that $t \ll 1$, and that the radius of the turnover curve remains small compared to the droplet radius, however the panels show the turnover curve at point 2 is close to $r = 0.75$. 

Both the value of the plate displacement $s(t)$ and the difference in the location of the fluid interfaces in the graphs and snapshots shown in Fig. \ref{fig:force_and_displacement_comparison} are small in comparison to the size of the droplet.  Hence, upon experimental observation, the physical system may not appear different to the stationary plate case. However the oscillations in the hydrodynamic force and the pressure differences in the snapshots show that flow inside the droplet is being significantly affected by the motion of the plate. This shows that just introducing substrate motion due to linear elasticity results in a substantial change in the dynamics of the droplet.

\subsection{Plate parameter comparisons}

In \textsection \ref{subsec:stationary_plate_comparison}, we showed in detail how the system behaves for specific values of the mass ratio $\alpha$, damping factor $\beta$ and stiffness factor $\gamma$. In the following we aim to study physical mechanisms represented by these parameters individually.

\begin{figure}[t]
    \centering
    \begin{subfigure}[t]{.32\textwidth}
        \centering
        \caption{$\beta = 0$,  $\gamma = 0$}
        \includegraphics[width=\textwidth]{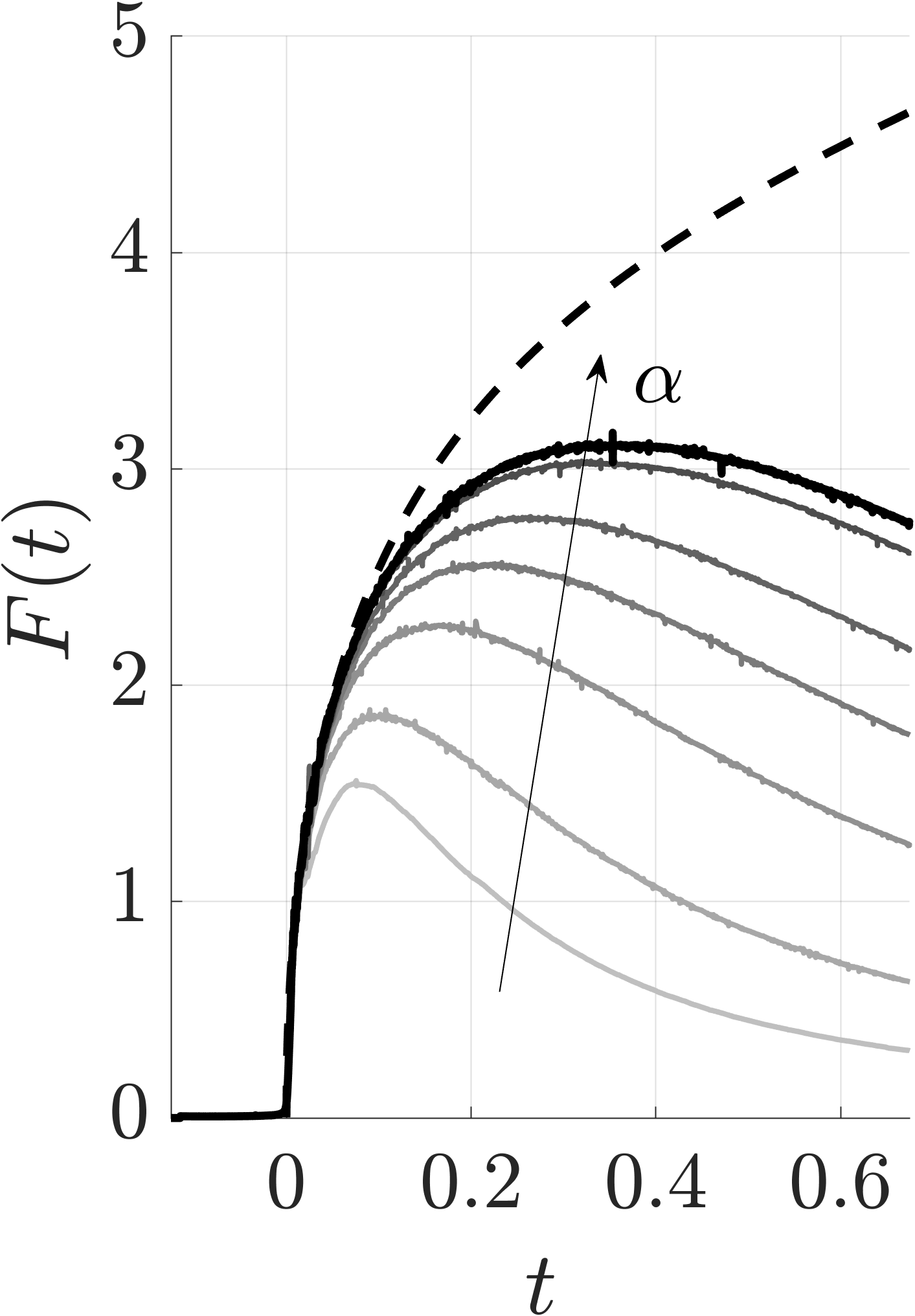}  
        \label{fig:alpha_comparison}
    \end{subfigure}
     \begin{subfigure}[t]{.32\textwidth}
        \centering
        \caption{$\alpha = 2$, $\gamma = 100$}
        \includegraphics[width=\textwidth]{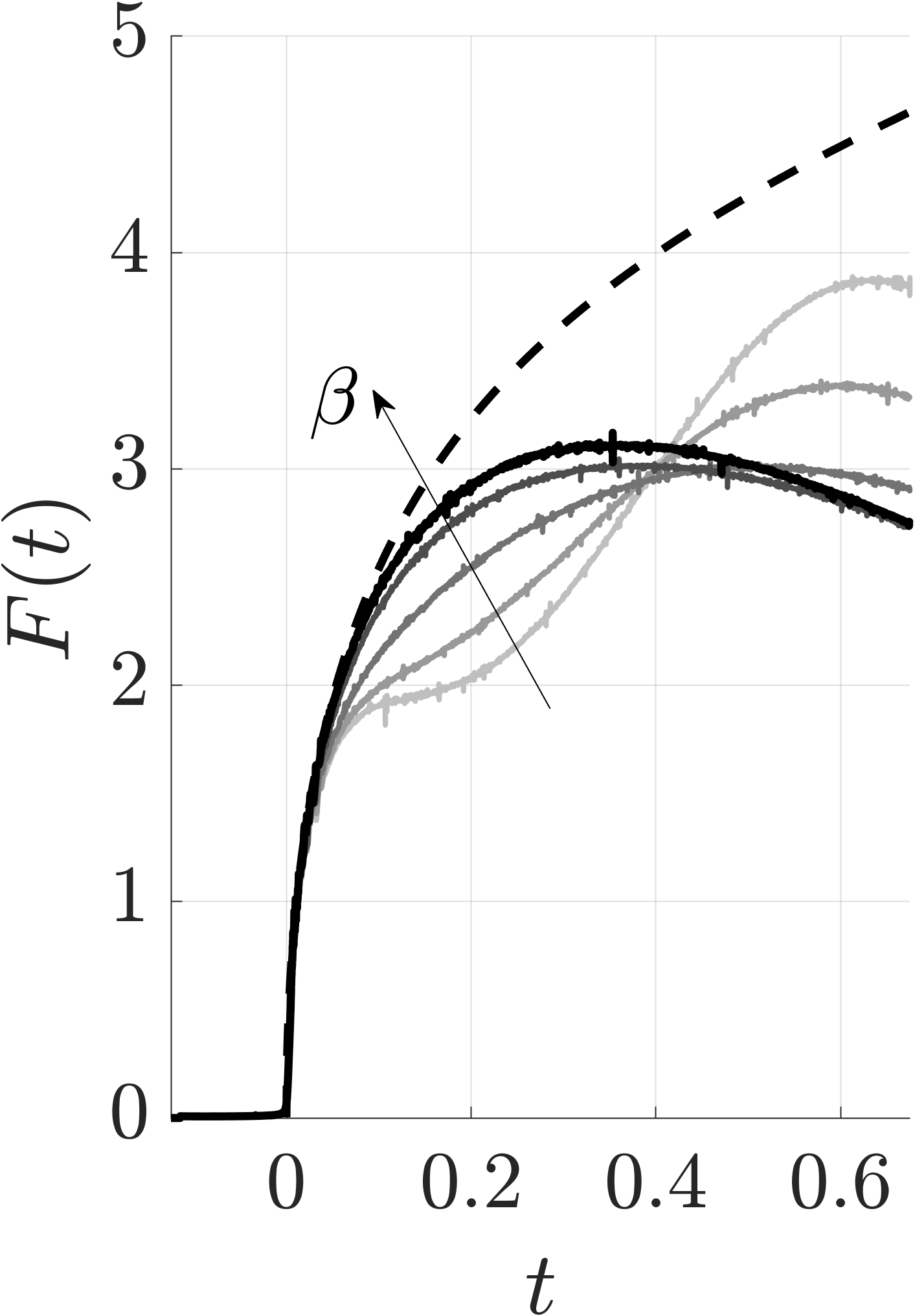} 
        \label{fig:beta_comparison}
    \end{subfigure}
     \begin{subfigure}[t]{.32\textwidth}
        \centering
        \caption{$\alpha = 2$, $\beta = 0$}
        \includegraphics[width=\textwidth]{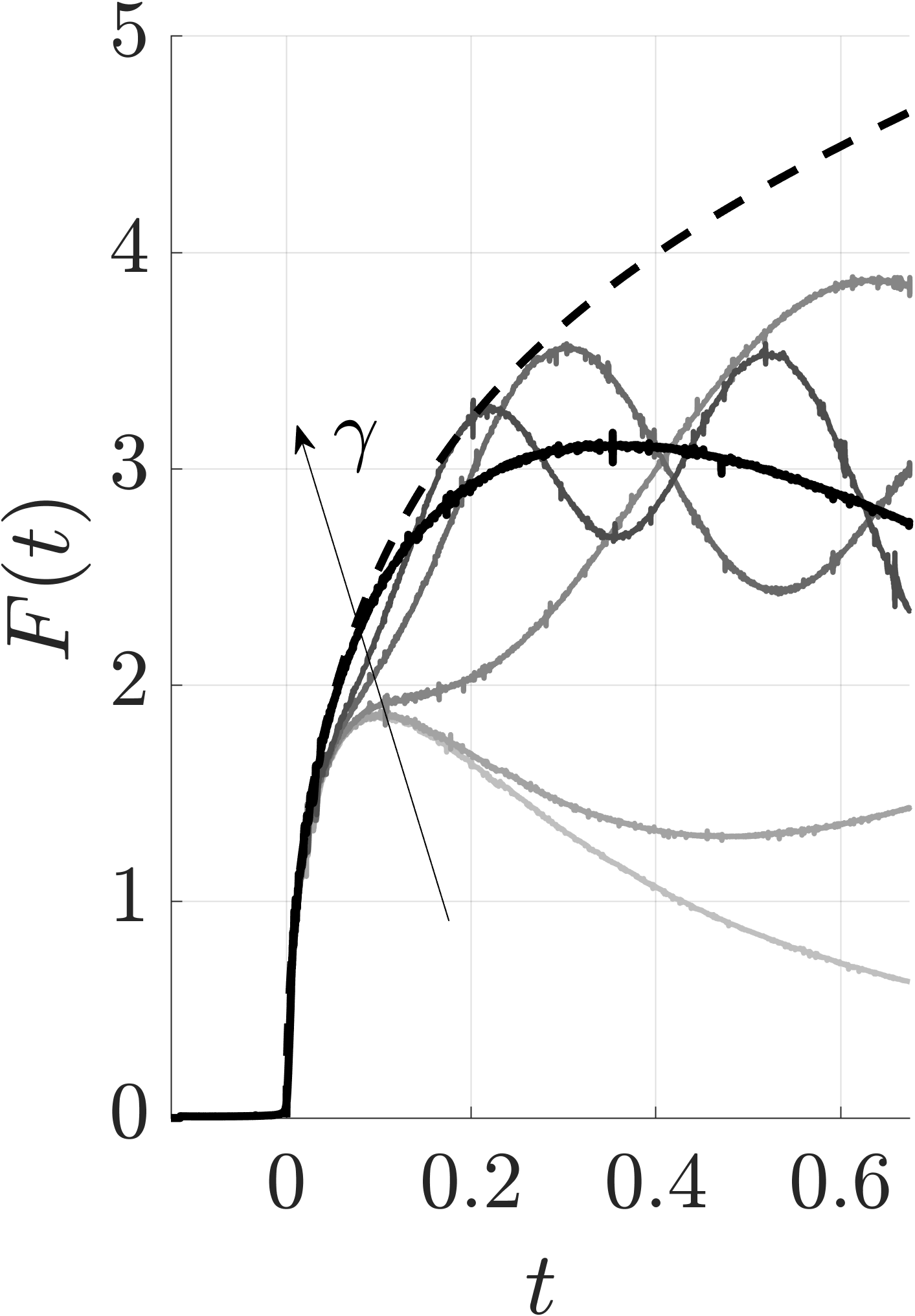} 
        \label{fig:gamma_comparison}
    \end{subfigure}
        \caption{Numerical values from the DNS for the hydrodynamic force on the plate $F(t)$, with solid grey lines showing moving plate cases with varying values of the mass ratio $\alpha$, damping factor $\beta$ and stiffness factor $\gamma$, and solid black lines showing the stationary plate case. The black dashed lines indicate the analytical solution to $F(t)$ for a stationary plate given by \eqref{eq:composite_force}. (a): $\beta$ = $\gamma$ = 0, and $\alpha$ ranges from 1 to  100. (b): $\alpha$ = 2, $\gamma = 100$ and $\beta = 0$, $0.25 \beta_c$, $\beta_c$ and $5 \beta_c$, where the critical damping value $\beta_c = 2 \sqrt{\alpha \gamma} \approx 28.28$. (c): $\alpha = 2$, $\beta = 0$ and $\gamma$ ranges from 0 to 1000.}
    \label{fig:parameter_comparisons}
\end{figure}

In order to systematically observe the effects of these physical mechanisms, we conducted a series of simulations for $-0.125 \leq t \leq 0.675$ with varying values of $\alpha$, $\beta$, $\gamma$. The hydrodynamic force $F(t)$ was calculated regularly and is shown by the solid greyscale lines in Fig. \ref{fig:parameter_comparisons}, where darker lines correspond to higher values of $\alpha$, $\beta$ or $\gamma$. For comparison, the solid black line and dashed line correspond to the numerical and analytical hydrodynamic force for the stationary plate case. Analytical solutions for the rest of the cases are not shown for visual clarity on the plots, however an analysis similar to that presented in \textsection \ref{subsec:stationary_plate_comparison} could be conducted for all each value of $\alpha$, $\beta$ and $\gamma$. 

The mass ratio $\alpha = M^* / \rho^*_l {R_d^*}^3$ represents (up to a constant) the ratio between the mass of the plate $M^*$ and the mass of the droplet $4 \pi \rho^*_l {R_d^*}^3 / 3$. Upon impact, the pressure of the droplet exerts a hydrodynamic force onto the plate, causing it to accelerate downwards. The downwards motion of the plate causes the pressure at the surface of the plate to decrease, in turn decreasing the hydrodynamic force. For lighter plates (smaller $\alpha$), this downwards motion will be faster, and hence we expect the hydrodynamic force will be lower in lighter plates than for heavier ones. The mass ratio $\alpha$ is varied from 1 to 100 in Fig. \ref{fig:alpha_comparison}, with $\beta = \gamma = 0$. In these cases, the only force acting on the plate is the hydrodynamic force of the droplet from above, hence the plate accelerates downwards at a rate depending on the mass ratio $\alpha$. It can be seen from Fig. \ref{fig:alpha_comparison} that increasing $\alpha$ causes the overall force to increase, tending towards the stationary plate value for large values of $\alpha$. In addition, the time at which the hydrodynamic force reaches a maximum increases as $\alpha$ increases, which happens once the plate has accelerated to its maximum velocity, resulting in a lower hydrodynamic force. It takes longer for this to happen the heavier the plate is, hence the time at which the maximum is reached increases as $\alpha$ increases.

The damping factor $\beta$ determines the amount of resistance to motion the dashpot exerts. We note that the ODE for the plate displacement \eqref{eq:dimensionless_plate_ode} under no external forcing has a critical damping value of $\beta = \beta_c = 2 \sqrt{\alpha \gamma}$.
If this unforced system were displaced from its equilibrium position and released, the undamped system ($\beta = 0$) would experience oscillations of a fixed amplitude about the equilibrium point. For $0 < \beta < \beta_c$ the system would be underdamped, and the amplitude of the oscillations would decay at a rate increasing with $\beta$. If $\beta > \beta_c$, the system would be overdamped and would exponentially return to equilibrium, returning more slowly with increasing $\beta$. Finally, if $\beta = \beta_c$, the system would be critically damped and would return to the equilibrium in the fastest time. However inclusion of the hydrodynamic forcing will alter these dynamics, and, in particular, we expect that higher values of $\beta$ would lead to smaller displacements from equilibrium due to the resistance to motion. In Fig. \ref{fig:beta_comparison}, the mass ratio and stiffness factor are fixed at $\alpha = 2$, $\gamma = 100$ such that the critical damping value is $\beta_c = 20 \sqrt{2} \approx 28.28$. The greyscale lines show the values of force for $\beta = 0$, $0.25 \beta_c$, $\beta_c$ and $5 \beta_c$. For $\beta = 0$, we can clearly see oscillations in the force, and the amplitude of these oscillations decreases when the system is underdamped for $\beta = 0.25 \beta_c$. These oscillations are suppressed in the case of critical damping $\beta = \beta_c$, where the force follows a trend that is initially lower than the stationary plate value, whereas the force approaches the stationary plate value for the overdamped case $\beta = 5 \beta_c$. Since the force depends predominantly on the hydrodynamic pressure in the droplet, the fact that the force follows the same behaviour of under-, over- and critical damping shows the strong influence the dashpot has on the behaviour of the droplet.

The strength of the elastic force from the compression of the spring is represented by the stiffness factor $\gamma$. In the absence of damping and external forcing, the solution of \eqref{eq:dimensionless_plate_ode} for $s(t)$ would oscillate with a dimensionless time period $T = 2 \pi \sqrt{\alpha / \gamma}$, hence an increase in the stiffness factor results in the oscillations having a shorter time period. The spring does work on the droplet via a vertical force equal to $\gamma s(t)$, so the droplet loses kinetic energy depending on how far below the $z$ axis the plate is displaced. Unlike damping, the elastic force is conservative, meaning the loss of kinetic energy from the droplet due to the elasticity is converted into potential energy in the spring, which is then in turn converted into kinetic energy via oscillations. Fig. \ref{fig:gamma_comparison} shows the hydrodynamic forces in systems for mass ratio and damping factor $\alpha = 2$ and $\beta = 0$, with stiffness factor $\gamma$ varying from 0 to 1000. As expected, we observe that the time periods of the oscillations decrease as $\gamma$ increases in Fig. \ref{fig:gamma_comparison}. For the two highest $\gamma$ values (500 and 1000), it can be seen that the values of $F(t)$ oscillate centred on the force value for the stationary plate case. This suggests that as $\gamma$ increases, although the frequency of oscillations increases, the amplitude of the oscillations would decrease and eventually the force would tend to the stationary plate value. 

Although the analytical solution was only shown for the stationary plate case in Fig. \ref{fig:parameter_comparisons}, it is worth noting the good agreement this solution has with the numerical values at early times for the majority of the moving plate cases shown. At early times, the velocity of the plate is still small, hence it does not significantly alter the hydrodynamic force. It is only once the plate has been accelerated that its motion affects the hydrodynamic force by doing work on the liquid. 

In this section, we have shown the rich variety in behaviours that the system exhibits as a result of the individual physical contributions due to the mass of the plate, strength of the dashpot and stiffness of the spring. These physical mechanisms result in previously unreported changes in the droplet dynamics, such as pressure oscillations which can be suppressed by energy losses due to damping. Although the magnitude of the plate displacement is small in comparison to the length scale of the droplet in all cases, the strong coupling observed between the plate displacement $s(t)$ and the hydrodynamic force $F(t)$ justify making use of models where the fluid-structure interaction is retained in order to accurately predict the dynamics of the droplet over these timescales.

\section{Summary and discussion}
\label{sec:conclusion}
In this paper we have presented two models for the vertical impact of a droplet onto a plate supported by a spring and a dashpot: an analytical model using matched asymptotic expansions and a full computational framework based on DNS. Although droplet impact onto elastic beams has been considered previously \cite{Pegg2018}, the analytical model we present is the first to consider the Wagner theory formulation where the substrate experiences both elastic forcing and damping. As opposed to previous axisymmetric models \cite{Oliver2002,Pegg2018}, we approximate the hydrodynamic force on the substrate using the leading-order composite expansion of the pressure between the outer and inner region (rather than just the outer region). Significantly, we found that the composite force shown in Fig. \ref{fig:force_and_displacement_comparison} is within 10\% of the numerical solution up to $t \approx 0.2$, in contrast to the force contribution due to the outer region only remains within 10\% of the numerical solution up to $t \approx 0.04$, which justifies considering the contributions from the inner region in order to extend the timescale in which such analytical models are valid more generally. Previous numerical investigations involving a plate-spring system \cite{Zhang2020} do not take into account forcing due to damping, and focus on the late-time dynamics of spreading and rebound, whereas we focus on the influence the plate motion has on the delicate early stages of impact in a high-speed context. 
Finally, the response of an elastic substrate on an impacting droplet has very recently been modelled using an effective boundary condition on the pressure in order to consider a stationary computational domain \cite{Howland2016a}. By contrast, in our model the substrate motion is resolved by using a moving frame of reference centred on the surface of the substrate, fully representing the fluid-structure interaction. 

The two proposed methodologies are distinct in their approach to understanding the system, and yet they are stronger in combination. In a problem with such violent topological changes over short scales, it is vital to have analytical results to both validate and inform our DNS platform. The analytical model provides guidance into key quantities, such as the location of the turnover point, which can be used as a prediction for the simulation duration and refinement strategy. In addition, the analytical model can be used to rapidly search for parameters where interesting coupling between the droplet and plate can be observed, rather than spending considerable computational resources searching for these regimes numerically. However, the analytical model relies on a series of assumptions, such as neglecting viscosity, surface tension and gravity, and is limited to early impact times. On its own, it is impossible to assess the consequence of these assumptions, and where they break down. By systematically comparing the analytical predictions to the numerical model, we can identify the regimes where these assumptions are valid and support the use of the analytical model as opposed to the costly DNS. If desired, the DNS can then be used to go beyond those regimes and study timescales inaccessible to the analytical model. It is only when used in conjunction that the predictive power and robustness of these models reach their full potential. 

Not only have the methods presented in this paper extended existing analytical and numerical models, they have also allowed us to provide physical insight into the dynamics of a novel, complex multi-phase system. We recognise that the displacement of the plate and perturbation of the free surface of the droplet are small, hence these models provide insight into a physical regime that would otherwise be very difficult to study experimentally. Through an extensive parameter study, we identified the influence that the mass ratio $\alpha$, damping factor $\beta$ and stiffness factor $\gamma$ have on the hydrodynamic force exerted by the droplet. In particular, we found that lighter plates (smaller $\alpha$) result in a lower value of the force; stiffer springs (higher $\gamma$) result in oscillations of higher frequency and that resistive dashpots (higher $\beta$) suppress the oscillations due to the elasticity from the spring.

The plate-spring-dashpot system is one of the simplest models for a flexible substrate, as it only allows for vertical motion. Droplet impact onto the end of a cantilever, such as a leaf, is more complex as the bending of the beam breaks the axisymmetry. However, as considered in \cite{Gart2015}, the deflection of the end of the beam can be modelled using the second order differential equation \eqref{eq:dimensionless_plate_ode} when the deflection is small (hence negligible bending). Therefore the model for the substrate motion considered in this paper could provide insight into the early time dynamics of droplet impact onto the end of cantilever beams. In addition, there is much scope to extend these models to more complex substrates, such as elastic membranes under tension, as previously studied experimentally \cite{Pepper2008}, further guided by recent analytical \cite{Pegg2018} and computational \cite{Patel2017,Sun2016} progress. In conclusion, we believe that the proposed mathematical framework embodies productive co-development and investigative interplay between rigorous state-of-the-art methodologies, providing a general and highly efficient route to studying complex systems involving fluid-structure interaction in the future.


%
\section*{Conflict of interest}

The authors declare that they have no conflict of interest.

\begin{appendices}

\section{Composite hydrodynamic force}
\label{appendix:composite_force}

For the analytical model, the hydrodynamic force is determined by integrating the pressure $p$ across the surface of the plate. As the leading-order solution in the outer region \eqref{eq:outer_pressure} diverges at the turnover point, the force contribution close to the turnover point must be an over-estimate. Hence the leading-order composite pressure between the outer and the inner region was found in \textsection \ref{subsec:composite_pressure} in order to determine the resulting composite force, where the force contribution from the splash sheet region was determined to be negligible. When viscosity is neglected, the dimensionless hydrodynamic force on the plate \eqref{eq:dimen_hydrodynamic_force} is given by
\begin{equation}
    F(t) = 2 \pi \int_0^{R_p} r p(r, - \epsilon^2 s(t), t) \dd{r}.
\end{equation}

The composite force is found by integrating the composite pressure \eqref{eq:composite_pressure} along the surface of the plate, which is determined by splitting the range of the integral across the respective asymptotic regions,
\begin{equation}
    F_\text{comp}(t) = F_\text{outer}(t) + F_\text{inner}(t) - F_\text{overlap}(t),
\end{equation}
where $F_\text{outer}(t)$ and $F_\text{inner}(t)$ are the result of integrating the leading-order outer pressure \eqref{eq:outer_pressure} and inner pressure \eqref{eq:inner_pressure} respectively, and $F_\text{overlap}(t)$ is the result of integrating the overlap pressure \eqref{eq:overlap_pressure}. The resulting force due to the outer region is hence,
\begin{equation}
\label{eq:outer_force}
\begin{aligned}
    F_\text{outer}(t) &= 2 \pi \int_0^{\epsilon d_0(t)} r \frac{1}{\epsilon} \hp_0(r / \epsilon, 0, t) \dd{r} = \frac{8}{9} \epsilon \int_0^{d_0(t)} \dv[2]{}{t} \left[\hr(d_0(t)^2 - \hr^2)^{3/2}\right] \dd{\hr} \\
    &= \frac{8}{9} \epsilon \dv[2]{}{t} \int_0^{d_0(t)} \hr (d_0(t)^2 - \hr^2)^{3/2} \dd{\hr} =\frac{8}{45} \epsilon \dv[2]{}{t} \left[d_0(t)^5\right] \\
    &= \frac{8}{9} \epsilon d_0(t)^3 (4\dot{d}_0(t)^2 + \ddot{d}_0(t)d_0(t)).
\end{aligned}
\end{equation}

For $F_\text{inner}(t)$, the integration variable is changed to the parameter $\eta$ from \eqref{eq:inner_pressure}. Then, $\eta_0(t)$ is defined such that $r = 0$ for $\eta = \eta_0(t)$, i.e.
\begin{equation}
    e^{2 \eta_0(t)} + 4 e^{\eta_0(t)} + 2 \eta_0(t) + 1 = \frac{\pi d_0(t)}{\epsilon^2 J(t)},
\end{equation}
where $\eta_0(t)$ must be solved for numerically for each $t$. Note that $\eta \to -\infty$ as $r \to \infty$, where $\tp_0$ decays exponentially. Hence we take the upper limit of the integral to be $r = \infty$, which introduces exponentially small errors. The resulting force due to the inner region is hence
\begin{equation}
\label{eq:inner_force}
\begin{aligned}
    F&_\text{inner}(t) = 2 \pi \int_0^\infty r \frac{1}{\epsilon^2} \tp_0(r / \epsilon^3 - d_0(t)/ \epsilon^2, 0, t) \dd{r} \\ 
    &= 2 \epsilon \pi \int_{-d_0(t)/\epsilon^2}^\infty (\epsilon d_0(t) + \epsilon^3 \tilr) \tp_0(\tilr, 0, t) \dd{\tilr} \\
    &= 8 \epsilon \dot{d}_0(t)^2 J(t) \int_{-\infty}^{\eta_0} \left(\epsilon d_0(t) - \epsilon^3 \frac{J(t)}{\pi} (e^{2 \eta} + 4 e^{\eta} + 2 \eta + 1)\right)e^\eta \dd{\eta} \\
    &= \frac{8 \epsilon^4 \dot{d}_0(t)^2 J(t)^2}{\pi}  e^{\eta_0(t)}\left[\frac{\pi d_0(t)}{\epsilon^2 J(t)} +1 - \frac{1}{3} e^{2 \eta_0(t)} - 2 e^{\eta_0(t)} - 2 \eta_0(t)\right].
\end{aligned}
\end{equation}

Finally the overlap force is 
\begin{equation}
\label{eq:overlap_force}
\begin{aligned}
    F_\text{overlap}(t) &= 2 \pi \frac{2\sqrt{2} d_0(t)^{3/2} \dot{d}_0(t)^2}{3 \pi \sqrt{\epsilon}} \int_0^{\epsilon d_0(t)} \frac{r}{\sqrt{\epsilon d_0(t) - r}} \dd{r} \\
    &= \frac{16 \sqrt{2}}{9} \epsilon d_0(t)^3 \dot{d}_0(t)^2.
\end{aligned}
\end{equation}

Combining \eqref{eq:outer_force}, \eqref{eq:inner_force} and \eqref{eq:overlap_force}, the composite force is
\begin{equation}
\begin{aligned}
    F_\text{comp}(t) &= \frac{8}{9} \epsilon d_0(t)^3 ((4- 2 \sqrt{2})\dot{d}_0(t)^2 + \ddot{d}_0(t)d_0(t)) \\
    &+ \frac{8 \epsilon^4 \dot{d}_0(t)^2 J(t)^2}{\pi}  e^{\eta_0(t)}\left[\frac{\pi d_0(t)}{\epsilon^2 J(t)} +1 - \frac{1}{3} e^{2 \eta_0(t)} - 2 e^{\eta_0(t)} - 2 \eta_0(t)\right],
\end{aligned}
\end{equation}
where $d_0(t)$ and $J(t)$ are given by \eqref{eq:d_solution} and \eqref{eq:J_solution} respectively. 

\end{appendices}

\bibliographystyle{spmpsci}      
\bibliography{references_edited.bib}   


\end{document}